\documentclass[twocolumn]{aastex61}
\usepackage{CJK} 

\usepackage{amsmath}
\usepackage{amssymb}
\usepackage{epsfig}
\usepackage{apjfonts}
\usepackage{natbib}
\usepackage{hyperref}
\usepackage{epstopdf}
\usepackage{verbatim}
\usepackage{enumitem}
\usepackage{color}
\setlist[enumerate]{itemsep=-1mm}
\usepackage{soul, xcolor}
\setstcolor{blue}

\usepackage{mathtools}

\usepackage{bigints}

\usepackage{tikz}

\makeatletter
\global\let\tikz@ensure@dollar@catcode=\relax
\makeatother

\received{Nov 5, 2020; Revised Feb 4, 2021; Accepted Feb 5, 2021}
\submitjournal{ApJ}

\begin{document}


\begin{CJK*}{UTF8}{gbsn}

\title{\large The Hawaii Infrared Parallax Program.\\V. New T-Dwarf Members and Candidate Members of Nearby Young Moving Groups}

\author[0000-0002-3726-4881]{Zhoujian Zhang (张周健)}
\affiliation{Institute for Astronomy, University of Hawaii at Manoa, Honolulu, HI 96822, USA}

\author[0000-0003-2232-7664]{Michael C. Liu}
\affiliation{Institute for Astronomy, University of Hawaii at Manoa, Honolulu, HI 96822, USA}

\author[0000-0003-0562-1511]{William M. J. Best}
\affiliation{The University of Texas at Austin, Department of Astronomy, 2515 Speedway, C1400, Austin, TX 78712, USA}

\author[0000-0001-9823-1445]{Trent J. Dupuy}
\affiliation{Gemini Observatory/NSF's NOIRLab, 670 N. A`ohoku Place, Hilo, HI, 96720, USA}
\affiliation{Institute for Astronomy, University of Edinburgh, Royal Observatory, Blackford Hill, Edinburgh, EH9 3HJ, UK}

\author[0000-0001-5016-3359]{Robert J. Siverd}
\affiliation{Gemini Observatory/NSF's NOIRLab, 670 N. A`ohoku Place, Hilo, HI, 96720, USA}

\begin{abstract}
We present a search for new planetary-mass members of nearby young moving groups (YMGs) using astrometry for 694 T and Y dwarfs, including 447 objects with parallaxes, mostly produced by recent large parallax programs from UKIRT and {\it Spitzer}. Using the BANYAN~$\Sigma$ and LACEwING algorithms, we identify 30 new candidate YMG members, with spectral types of T0$-$T9 and distances of $10-43$~pc. Some candidates have unusually red colors and/or faint absolute magnitudes compared to field dwarfs with similar spectral types, providing supporting evidence for their youth, including 4 early-T dwarfs. We establish one of these, the variable T1.5 dwarf 2MASS~J21392676+0220226, as a new planetary-mass member ($14.6^{+3.2}_{-1.6}$~M$_{\rm Jup}$) of the Carina-Near group ($200\pm50$~Myr) based on its full six-dimensional kinematics, including a new parallax measurement from CFHT. The high-amplitude variability of this object is suggestive of a young age, given the coexistence of variability and youth seen in previously known YMG T~dwarfs. Our four latest-type (T8$-$T9) YMG candidates, WISE~J$031624.35+430709.1$, ULAS~J$130217.21+130851.2$, WISEPC~J$225540.74-311841.8$, and WISE~J$233226.49-432510.6$, if confirmed, will be the first free-floating planets ($\approx2-6$~M$_{\rm Jup}$) whose ages and luminosities are compatible with both hot-start and cold-start evolutionary models, and thus overlap the properties of the directly-imaged planet 51~Eri~b. Several of our early/mid-T candidates have peculiar near-infrared spectra, indicative of heterogenous photospheres or unresolved binarity. Radial velocity measurements needed for final membership assessment for most of our candidates await upcoming 20--30~meter class telescopes. In addition, we compile all 15 known T7$-$Y1 benchmarks and derive a homogeneous set of their effective temperatures, surface gravities, radii, and masses. 
\end{abstract}

\section{Introduction}
\label{sec:introduction}
A plethora of planetary-mass objects have been discovered beyond our solar system in the past 25 years. Among these objects, gas-giant planets ($\approx 1-13$~M$_{\rm Jup}$) that are either wide-separation companions to stars or brown dwarfs \citep[e.g.,][]{2010MNRAS.405.1140G, 2014ApJ...787....5N, 2017AJ....154..262M, 2018AJ....156...57D} or free-floating objects \citep[e.g.,][]{2013ApJ...777L..20L, 2017ApJ...837...95B, 2017AJ....153..196S, 2018ApJ...858...41Z} are a valuable subset for high-quality emission spectroscopy, given the lack of the contaminating light from host stars. These objects thereby serve as excellent laboratories to study self-luminous exoplanet atmospheres, as well as exoplanet formation and evolution. As they are too low in mass to fuse either hydrogen or deuterium in their cores, planetary-mass objects contract, cool, and fade after their initial formation \citep[e.g.,][]{2001RvMP...73..719B, 2007ApJ...655..541M}. Consequently, searches for self-luminous giant planets have focused on the nearest young ($\approx 10-200$~Myr) moving groups (YMGs) and stellar associations, where planetary-mass objects are bright enough to be directly detected. Moreover, by virtue of their shared membership, these planetary-mass objects can adopt the age estimates inferred for the stellar members of the same groups, making them ``age benchmarks'' \citep[e.g.,][]{2006MNRAS.368.1281P, 2007ApJ...660.1507L} for testing models of substellar evolution and ultracool atmospheres.

Substantial progress has been made to identify new members of nearby YMGs and has spawned a variety of methods for membership assessment \citep[e.g.,][]{2004ApJ...613L..65Z, 2005ApJ...634.1385M, 2006A&A...460..695T, 2012ApJ...758...56S, 2013ApJ...762...88M, 2017AJ....153...18B, 2019ApJ...877...60B, 2017AJ....153...95R, 2018ApJ...856...23G, 2019MNRAS.489.3625C}. These methods rely on the objects' space motions to establish membership, along with spectrophotometric evidence to establish their youthfulness. With precise proper motions and parallaxes, {\it Gaia}~DR2 \citep[][]{2016AandA...595A...1G, 2018AandA...616A...1G} has enabled kinematic studies of the solar neighborhood and greatly expanded the stellar and substellar census of nearby associations \citep[e.g.,][]{2018ApJ...862..138G}. However, optical data from {\it Gaia} are not sensitive to planetary-mass objects, whose spectral energy distributions peak at longer wavelengths. Deep optical and near-infrared sky surveys are valuable resources to find free-floating planets, including Pan-STARRS1 \citep[PS1;][]{2016arXiv161205560C}, UKIDSS \citep[][]{2007MNRAS.379.1599L, 2012yCat.2314....0L}, and the {\it WISE} surveys \citep[][]{2010AJ....140.1868W, 2014yCat.2328....0C, 2020ApJS..247...69E, 2020arXiv201213084M}. These catalogs provide proper motions, but parallaxes are either lacking or low-accuracy given the limited number of epochs and time baseline, thus inhibiting the identification of new association members. 

Given these challenges, the current planetary-mass census of nearby associations is largely incomplete and has a significant deficit at T and Y spectral types. Mid- to late-T dwarfs are among the most common field population in the solar neighborhood \citep[e.g.,][]{2012ApJ...753..156K, 2013MNRAS.433..457B, 2015MNRAS.449.3651M, 2021AJ....161...42B}, but we still have limited census of such objects at young ages. Only a handful of T-dwarf (and no Y-dwarf) YMG members have been found to date \citep{2014ApJ...787....5N, 2015Sci...350...64M, 2015ApJ...808L..20G, 2017ApJ...841L...1G, 2018ApJ...854L..27G}, and a larger sample of such objects is needed to investigate their atmospheres ($T_{\rm eff} \approx 500 - 1200$~K) at low surface gravities. 

The recent completion of infrared parallax programs by \cite{2019ApJS..240...19K} and by \cite{2020AJ....159..257B} has provided the largest batch (over 300 objects) of new proper motions and parallaxes for LTY-type ultracool dwarfs, and these precise data open the door to a large-scale search for late-type YMG members. In this work, we combine available astrometry and radial velocities of 694 T and Y dwarfs (447 objects with parallaxes) to identify new and candidate members of nearby YMGs (Section~\ref{sec:identification}). By studying the astrometric, photometric, and spectroscopic properties of our candidates, we have confirmed a new planetary-mass member in the Carina-Near moving group and found 29 other T-dwarf candidate members, including several with unusual spectrophotometric properties (Section~\ref{sec:property}). Finally, we provide a summary and discuss future follow-up of our candidates (Section~\ref{sec:summary}).

\begin{figure*}[t]
\begin{center}
\includegraphics[height=6.5in]{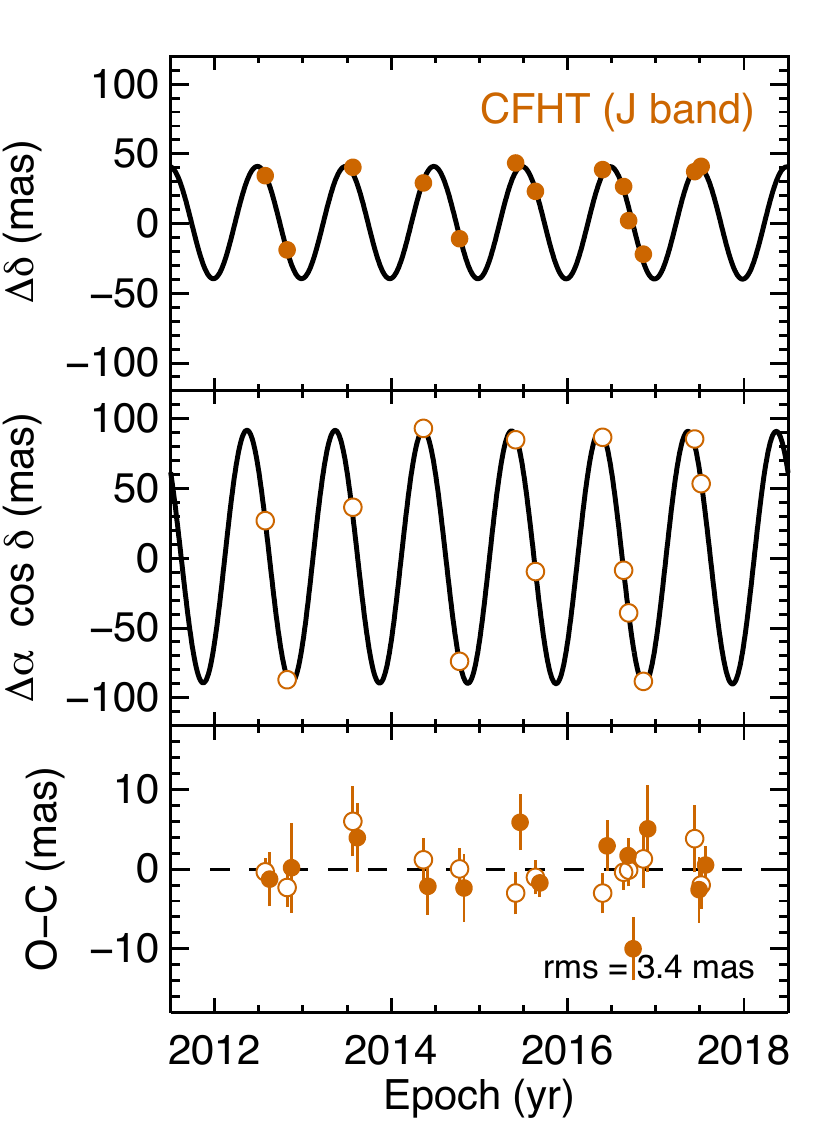}
\caption{Relative astrometry of 2MASS~J$2139+0220$ from CFHT/WIRCam. In the top two panels, the best-fit proper motions have been subtracted (for display purposes only), and the best-fit parallax solution is plotted as a solid black line. The bottom panel shows the residuals about the best-fit solution, with small x-axis offsets added to the $\Delta\delta$ residuals to more clearly show error bars. Our best-fit solution has a reduced $\chi^{2} = 1.04$ with 21 degrees of freedom.}
\label{fig:cfht_plx_2139}
\end{center}
\end{figure*}

\section{Identification of New Young Moving Group Members}
\label{sec:identification}

\subsection{Data}
\label{subsec:plx_pm_rv}
We start our analysis using The UltracoolSheet\footnote{\url{http://bit.ly/UltracoolSheet}} \citep[][]{ultracoolsheet}, a catalog of astrometry, photometry, spectroscopy, and multiplicity for over 3,000 ultracool dwarfs and imaged exoplanets. Developed from compilations of ultracool dwarfs by \cite{2012ApJS..201...19D}, \cite{2013Sci...341.1492D}, \cite{2016ApJ...833...96L}, \cite{2018ApJS..234....1B}, and \cite{2021AJ....161...42B}, The UltracoolSheet is complete for all spectroscopically confirmed objects with $\geqslant$L0 spectral types known prior to April 15, 2015 and is further augmented by new ultracool dwarfs discovered by \cite{2015ApJ...814..118B, 2017ApJ...837...95B} and all imaged exoplanets discovered since then. Sky positions, proper motions, parallaxes, and radial velocities of objects in The UltracoolSheet are compiled from numerous catalogs, including {\it Gaia}~DR2 \citep[][]{2016AandA...595A...1G, 2018AandA...616A...1G}, PS1 \citep[][]{2016arXiv161205560C}, UKIDSS \citep[][]{2007MNRAS.379.1599L, 2012yCat.2314....0L}, AllWISE \citep[][]{2014yCat.2328....0C}, the SIMBAD Astronomical Database \citep[][]{2000A&AS..143....9W}, and recent large near-infrared parallax programs by \cite{2017AJ....154..147D}, \cite{2018MNRAS.481.3548S}, \cite{2019ApJS..240...19K}, and \cite{2020AJ....159..257B}. 

For one particular T dwarf, 2MASS~J$21392676+0220226$ (2MASS~J$2139+0220$), we also include our new astrometry measurements. We monitored this object with the facility infrared camera WIRCam \citep{2004SPIE.5492..978P} on the Canada-France-Hawaii Telescope (CFHT) from 2012--2017. Using 5-second exposures in the $J$ band, we achieved signal-to-noise ratios (S/Ns) of $90-160$ on the target in individual frames, obtaining an average of 18 frames per epoch, from which we measured the $(x,y)$ positions of it and 126 reference stars. Using our custom pipeline \citep{2012ApJS..201...19D, 2015ApJ...805...56D}, we reduced these individual measurements into high-precision multi-epoch relative astrometry, with the absolute calibration provided by 91 low-proper-motion 2MASS stars \citep{2003tmc..book.....C}. We derived the relative parallax and proper motion for 2MASS~J2139+0220 using our standard MCMC approach and then, in order to be consistent with our many previously published CFHT parallaxes, converted to an absolute reference frame using the Besan\c{c}on galaxy model to simulate the distances of the reference stars \citep{2003A&A...409..523R}. Our thirteen epochs of astrometry spanning 4.94~years yield an absolute parallax of $96.5 \pm 1.1$~mas and proper motion of $(489.7\pm0.7,125.0\pm0.8)$~mas~yr$^{-1}$. We found a very reasonable reduced $\chi^2 = 1.04$ for our best-fit solution with 21~degrees of freedom (Figure~\ref{fig:cfht_plx_2139}).

Multiple astrometric and kinematic measurements for the same objects in The UltracoolSheet are unified following the approach described below (see \citealt{ultracoolsheet} for more details). For objects that are companions in binary systems, we assume the companion has the same sky position, proper motion, and parallax as its host stars if (1)~this binary system has an angular separation of $\leqslant 1''$, or (2)~the companion has no direct astrometry from the existing catalogs. We also adopt the host stars' radial velocities if these values are lacking or have lower precision for the companions.

We computed J2000 sky positions of all ultracool dwarfs using the following preferences (from highest to lowest): {\it Gaia}~DR2, PS1, UKIDSS, AllWISE, and SIMBAD. The coordinates of PS1 and UKIDSS are given at the observed epoch, so we computed J2000 coordinates using the reported epochs and proper motions. Such calculation is also performed by SIMBAD for {\it Gaia}~DR2.

The final adopted proper motions and parallaxes of ultracool dwarfs are taken from {\it Gaia}~DR2 if available and otherwise from the most precise measurements among PS1 and the literature. The objects' PS1 parallaxes are required to have the S/N of at least $5$. We allow the adopted proper motions and parallaxes of a given object to come from different references. The radial velocities are mostly obtained from SIMBAD, available for nearly 1,000 objects. 

We identify new members and candidate members of nearby YMGs by using the resulting compilation of astrometry and radial velocities of 694 T and Y dwarfs in The UltracoolSheet (447 objects with parallaxes), including both single objects and components of resolved binary/multiple systems. For objects without trigonometric parallaxes, we use the photometric distances available in The UltracoolSheet, with final values calculated from $W2$, $K$, or $J$ band\footnote{We adopt the objects' $W2$-based distances if their {\it WISE} photometry exists and is not contaminated by nearby sources (i.e., ``nb == 1''). Otherwise, we adopt photometric distances computed from $K_{\rm 2MASS}$ (more preferred) or $K_{\rm MKO}$ band for $<$T4.5 dwarfs, and from $J_{\rm 2MASS}$ (more preferred) or $J_{\rm MKO}$ band for later-type objects.}. Photometric distances are computed using the \cite{2012ApJS..201...19D} relation between absolute magnitudes and spectral types established for field-age, high-gravity objects. This relation differs for young, low-gravity objects, especially at the L/T transition \citep[e.g.,][]{2013ApJ...777L..20L, 2016ApJ...833...96L, 2016ApJS..225...10F, 2020ApJ...891..171Z}, so we treat with caution candidate YMG members identified using photometric distances.

\subsection{Membership Assessment}
\label{subsec:membership}
We use both BANYAN~$\Sigma$ \citep[version~1.2; ][]{2018ApJ...856...23G} and LACEwING \citep[][]{2017AJ....153...95R} to evaluate whether a given object in The UltracoolSheet is a YMG member. BANYAN $\Sigma$ is a Bayesian inference framework that compares an object's sky position and proper motion (as well as its parallax and radial velocity when available) to those of bona fide members of 29 young moving groups and associations ($\approx 1-800$~Myr) within 150~pc and field stars simulated by the Besan\c{c}on Galactic model \citep{2003A&A...409..523R}, and then computes a membership probability based on the object's Galactic coordinates and space velocity ($XYZUVW$). A threshold value for the computed Bayesian probabilities is needed to assess the objects' membership and the robustness of such a threshold can be tested against known YMG members and synthetic field stars, with the results described by the confusion matrix and derived quantities, including the true-positive rate (i.e., the fraction of known members recovered) and false-positive rate (i.e., the fraction of contaminating field stars that are incorrectly classified as members). In principle, different probability thresholds are needed for different associations in order to achieve the same recovery/contamination rate, given that the YMGs have a variety of sizes, distances, and membership completeness. To reduce such association dependence for the threshold, \cite{2018ApJ...856...23G} customized their Bayesian priors and designed BANYAN $\Sigma$ to produce similar recovery rates\footnote{With the $90\%$ probability threshold, BANYAN~$\Sigma$ can recover $50\%$ (proper motion only), $68\%$ (proper motion and radial velocity), $82\%$ (proper motion and parallax), and $90\%$ (proper motion, radial velocity, and parallax) of bona fide members of all associations \citep{2018ApJ...856...23G}. The false-positive rates for different associations depend on their angular sizes and characteristic kinematics but are usually $\leqslant10^{-3}$.} for all 29 YMGs at a $90\%$ threshold value. Therefore, the $90\%$ probability reported by BANYAN~$\Sigma$ is not a metric for the true membership, but rather a value chosen to allow the classification performance among different YMGs to converge \citep[see Section~7 of][]{2018ApJ...856...23G}.

LACEwING is a frequentist inference framework that compares an object's kinematics with those of nearby YMGs using the observed quantities (sky position, proper motion, parallax, radial velocity) rather than $XYZUVW$ as done by BANYAN~$\Sigma$. LACEwING incorporates 16 young moving groups and associations ($\approx 5-800$~Myr) within 100~pc, which are all included in BANYAN~$\Sigma$ but with slightly different lists of bona fide members. The resulting LACEwING probability directly describes the likelihood that a given object is a kinematic member in each YMG, and the probabilities among all YMGs do not necessarily add up to $100\%$ by design. \cite{2017AJ....153...95R} suggested probability thresholds of $66\%$, $40\%$, and $20\%$ to select high-, moderate-, and low-probability candidate members, respectively. Unlike BANYAN~$\Sigma$, a given probability threshold does not promise the similar recovery rate or contamination rate among different YMGs. Also, within the same YMG, any (positive) probability threshold can lead to a very wide range of recovery rates (spanning $0\%-100\%$) depending on the number and type of input astrometric parameters \citep[e.g., see Figure~4 in][]{2017AJ....153...95R}.

\begin{figure*}[t]
\begin{center}
\includegraphics[height=6.4in]{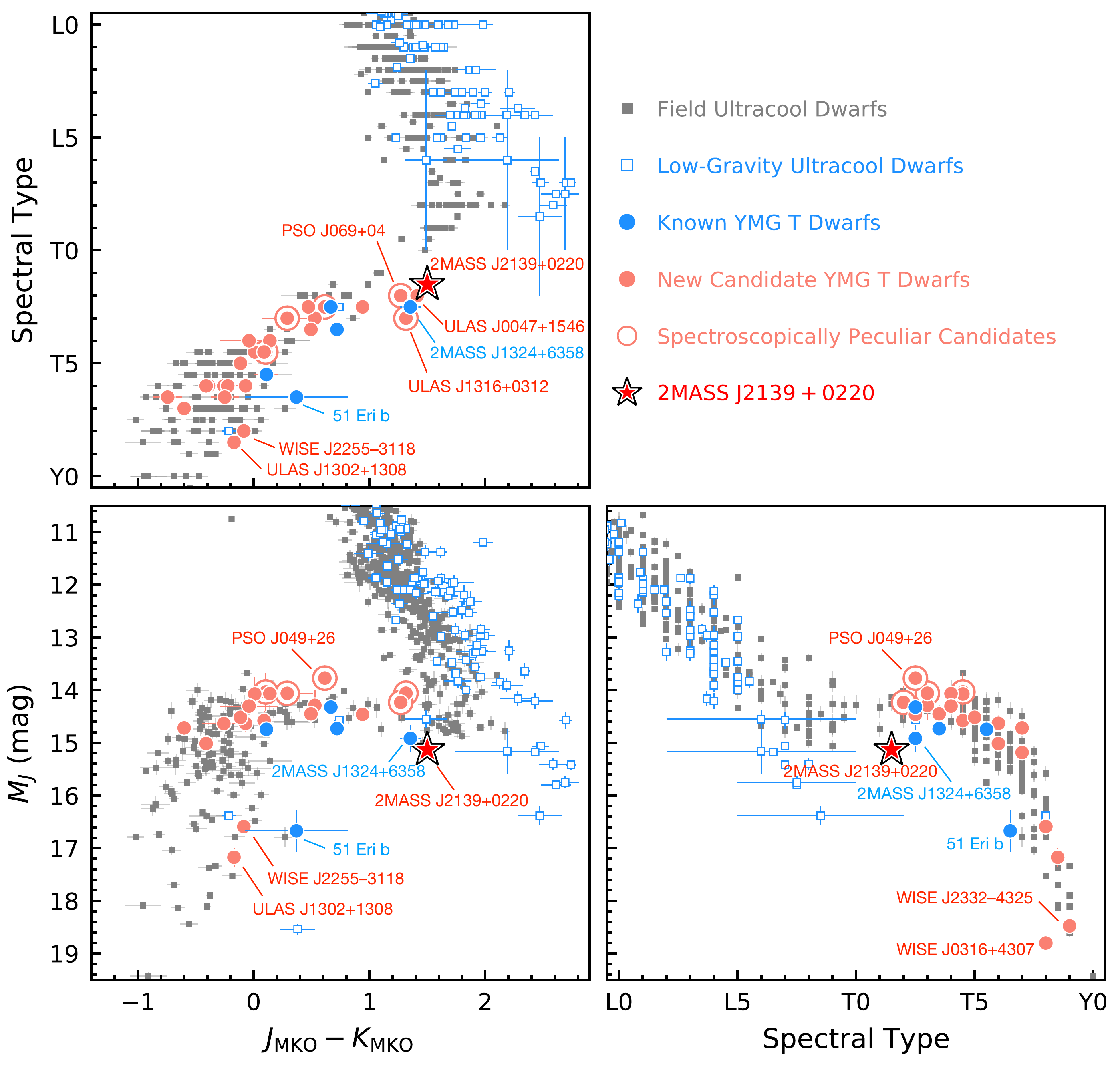}
\caption{Near-infrared photometry of our newly confirmed Carina-Near member 2MASS~J$2139+0220$ (red star), our newly identified YMG candidate members (light red circles), and recovered previously known T-dwarf YMG members (blue circles). We use open red circles to mark 7 YMG candidates with peculiar spectra indicative of either atmospheric variability or unresolved binarity (Section~\ref{subsec:spec}). Gray squares show known field dwarfs from The UltracoolSheet that have absolute magnitudes with S/N$>$5 and are not young, resolved binaries, or subdwarfs. Blue open squares show low-gravity L and T dwarfs in the field or YMGs.  }
\label{fig:cmd}
\end{center}
\end{figure*}

We feed sky positions, proper motions, parallaxes, and radial velocities of The UltracoolSheet objects into BANYAN $\Sigma$ and LACEwING, and then select T and Y dwarfs with $\geqslant 80\%$ BANYAN~$\Sigma$ or $\geqslant 66\%$ LACEwING membership probabilities. For BANYAN~$\Sigma$, our chosen threshold will lead to a higher recovery rate of known YMG members than using $90\%$, with the specific enhancement depending on associations \cite[see Figure~12 of][]{2018ApJ...856...23G}, but false-positive rates will increase as well. Using the BANYAN~$\Sigma$ and LACEwING results, we recover all the 5 T-dwarf YMG members known to date and find 30 new T-type candidate members\footnote{One of our candidate members, CFHT-Hy-20 (T2.5), was previously suggested as a Hyades member by \cite{2008AandA...481..661B} based on the photometry and proper motion. The more precise proper motion, as well as the new parallax, measured by \cite{2016ApJ...833...96L} supported the object's Hyades membership. Here we also identify this object as a candidate, but do not consider it to be confirmed, given that a radial velocity measurement is lacking. In addition, while this paper was under review, new astrometry of 5 YMG candidates and 1 previously known YMG member became available from \cite{2020arXiv201111616K}, which does not alter these objects' candidacy but does lead to slightly different membership probabilities (see footnote~$b$ of Table~\ref{tab:members}). Also, two other objects (WISE~J033651.90$+$282628.8 and PSO~J319.3102$-$29.6682) were previously considered as candidates but are now excluded from our analysis, since their YMG membership probabilities do not pass our criteria by using these objects' new and more precise astrometry from \cite{2020arXiv201111616K}.} (Table~\ref{tab:members}). In the following section, we study the photometric, spectroscopic, and physical properties of our candidates and discuss their membership.

\section{Properties of Candidate Members}
\label{sec:property}

\subsection{Photometric Properties}
\label{subsec:phot}
Figure~\ref{fig:cmd} and Table~\ref{tab:phot} present near-infrared photometry of our YMG candidates. Several early-T dwarfs exhibit $\approx 0.8$~mag redder $J-K$ colors than field dwarfs with similar spectral types, including four of our candidate members, 2MASS~J$2139+0220$ (T1.5), ULAS~J$004757.41+154641.4$ (ULAS~J$0047+1546$; T2), PSO~J$069.7303+04.3834$ (PSO~J$069+04$; T2), and ULAS~J$131610.13+031205.5$ (ULAS~J$1316+0312$; T3), as well as one previously known AB~Doradus member identified by \cite{2018ApJ...854L..27G}, 2MASS~J$13243553+6358281$ (2MASS~J$1324+6358$; T2.5). Also, both 2MASS~J$2139+0220$ and 2MASS~J$1324+6358$ have $\approx 0.6$~mag fainter $J$-band absolute magnitudes than the field sequence. The anomalous photometry of these objects provides evidence for their youth, given that the L/T transition of ultracool dwarfs is surface-gravity dependent, with young, lower-gravity objects having fainter, redder near-infrared photometry than their older, higher-gravity counterparts at the same spectral type \citep[e.g.,][]{2006ApJ...651.1166M, 2011ApJ...733...65B, 2016ApJS..225...10F, 2016ApJ...833...96L}. 

The gravity dependence of the L/T transition likely diminishes from early to later T types \citep{2020ApJ...891..171Z}. Therefore the membership of our remaining candidates, whose photometry follows the field sequence, is still plausible. Despite this, two of our late-T candidate members, WISEPC~J$225540.74-311841.8$ (WISE~J$2255-3118$; T8) and ULAS~J$130217.21+130851.2$ (ULAS~J$1302+1308$; T8.5), as well as one previously known $\beta$~Pictoris member, 51~Eri~b \citep[T6.5;][]{2015Sci...350...64M}, have redder $J-K$ colors than field dwarfs by $0.4-0.8$~mag. Also, both WISE~J$031624.35+430709.1$ (WISE~J$0316+4307$; T8) and 51~Eri~b have fainter $J$-band absolute magnitudes than the field sequence by $1.6-2.2$~mag. These three late-T candidates are therefore likely young as well given their similar photometry to 51~Eri~b.

\begin{figure*}[t]
\begin{center}
\includegraphics[height=1.6in]{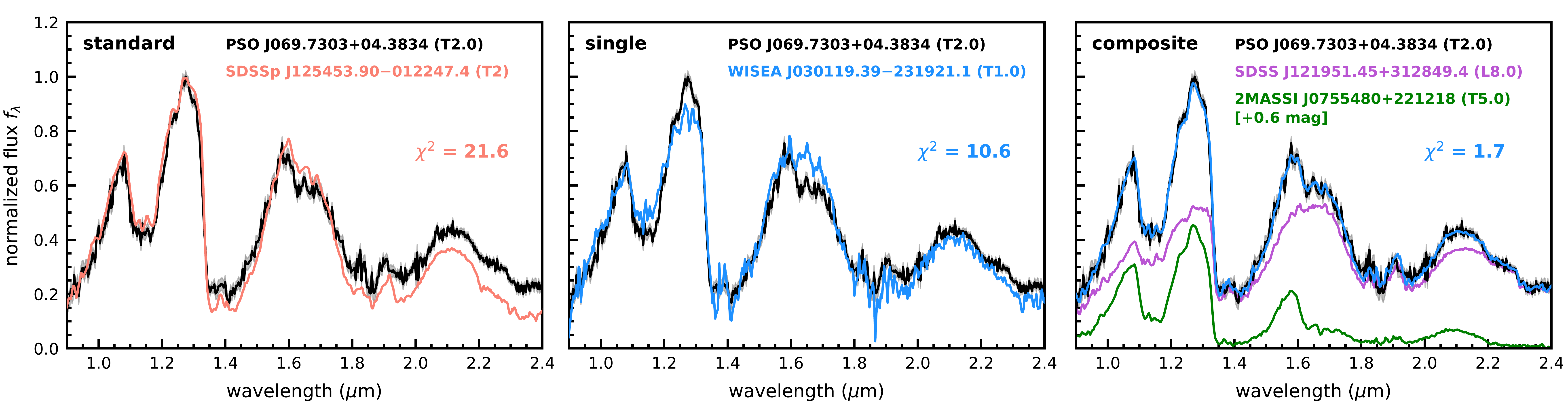}
\includegraphics[height=1.6in]{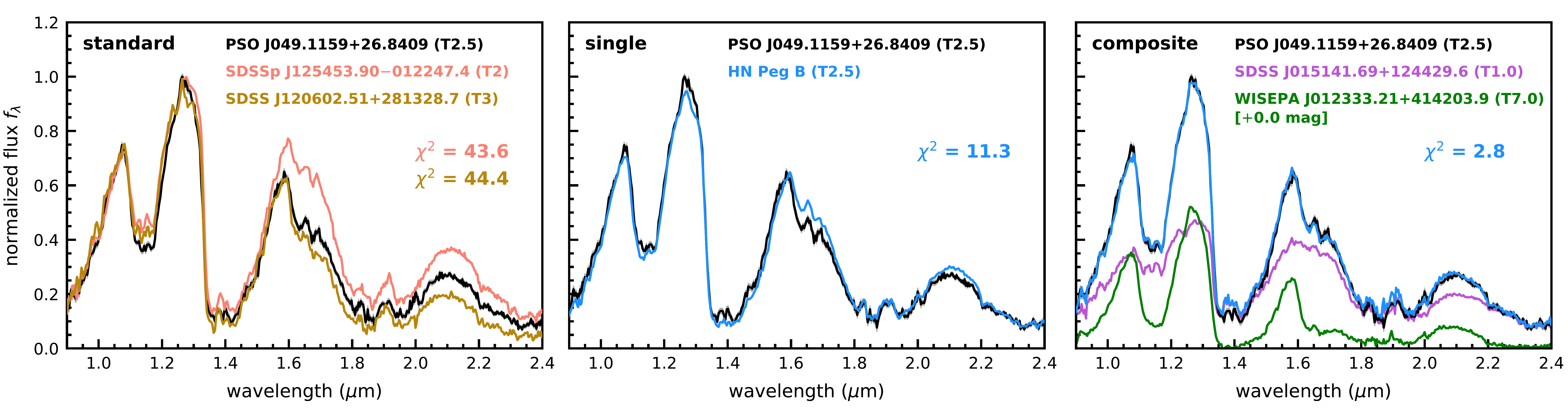}
\includegraphics[height=1.6in]{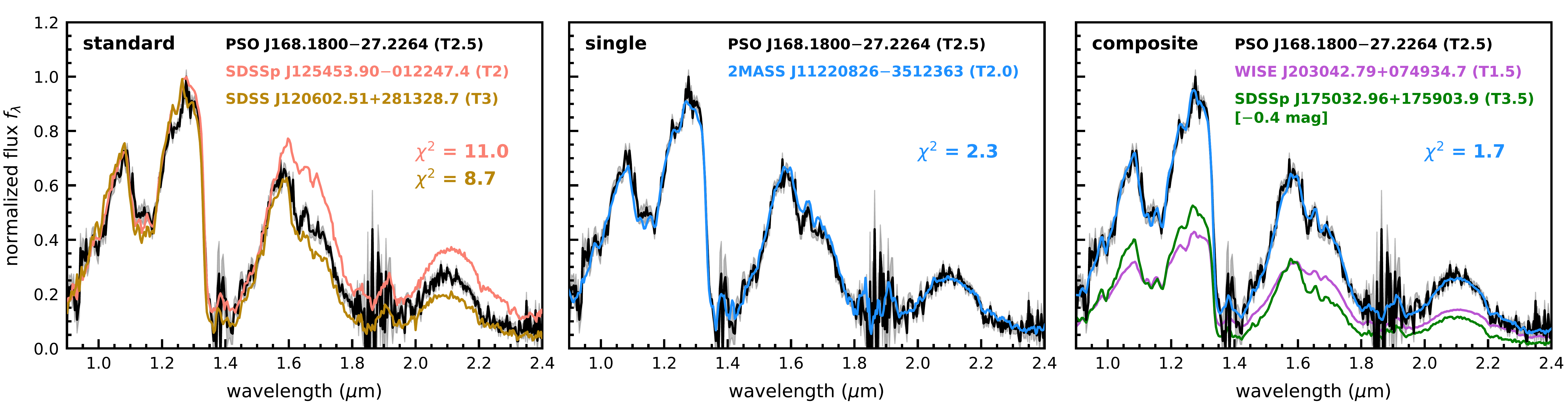}
\includegraphics[height=1.6in]{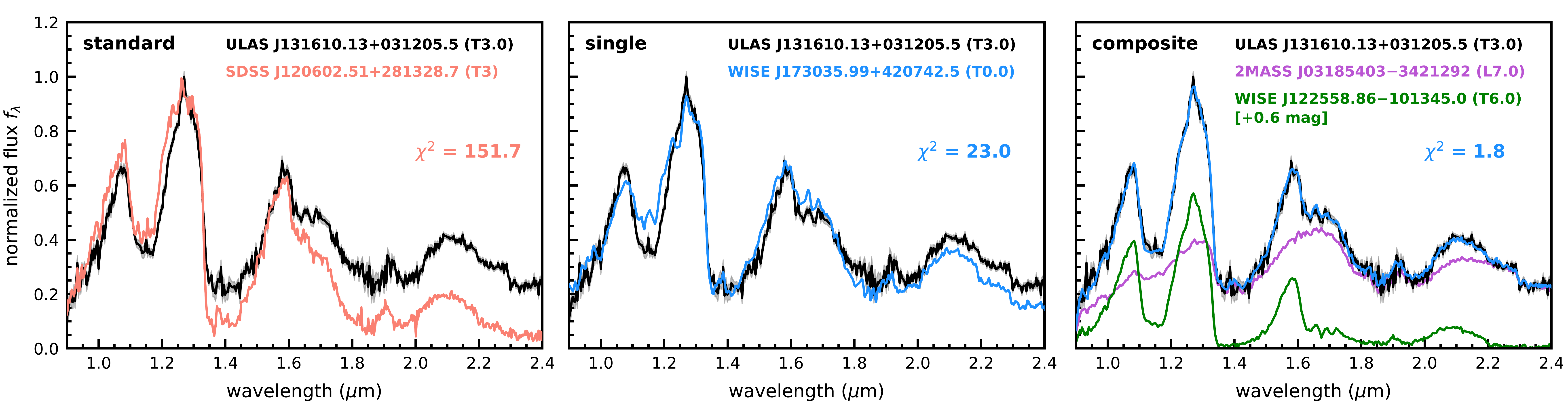}
\caption{Near-infrared spectra of our 4 strong composite candidates, compared with spectral standards (left), the best-fit single-object spectral template (middle), and the best-fit composite spectral template (right), with $\chi^{2}$ values labeled. In the left panel for each object, we show only one spectral standard if the object has an integer spectral type and two standards for objects with half types. In the right panel, we use purple and green for the primary and the secondary components, respectively, with the composite spectra shown in blue. The value in the brackets (green) indicates the magnitude offset we have added to the absolute $K_{\rm MKO}$ of the secondary when flux-calibrating its spectrum and generating the composite spectral template. }
\label{fig:spec_strong}
\end{center}
\end{figure*}

\subsection{Spectroscopic Properties}
\label{subsec:spec}
Among our 30 candidate YMG members, 20 objects have low-resolution ($R \sim 100$) near-infrared ($0.8-2.5$~$\mu$m) spectra observed by the NASA Infrared Telescope Facility (IRTF) with the facility spectrograph SpeX \citep[][]{2003PASP..115..362R} in prism mode. Here we investigate if any candidates exhibit spectral peculiarity, which is indicative of atmospheric variability or unresolved binarity, with the former related to the rotation of ultracool dwarfs with inhomogeneous photospheric condensate clouds \citep[e.g.,][]{2014ApJ...793...75R} and/or temperature fluctuations \citep[e.g.,][]{2020A&A...643A..23T}. In both the variability and binary scenarios, a peculiar spectrum might be described by the composite of two (parts of) photospheres with different effective temperatures. Therefore, empirical spectral indices designed to identify unresolved binaries \citep[e.g.,][]{2010ApJ...710.1142B, 2014ApJ...794..143B} can also find objects with high-amplitude photometric variability \citep[e.g.,][]{2012ApJ...750..105R, 2013AJ....145...71K, 2015ApJ...801..104H, 2016ApJ...826....8Y, 2019AJ....157..101M}. 

We have visually compared the IRTF/SpeX spectra of our 20 candidates to spectral standards from \cite{2006ApJ...637.1067B} and \cite{2011ApJ...743...50C} to identify any spectral peculiarity. We have also computed the \cite{2010ApJ...710.1142B} quantitative spectral indices to identify objects with spectra indicative of composite photospheres or unresolved binarity. As a result, we find 4 ``strong'' composite candidates, meeting at least 3 out of 6 \cite{2010ApJ...710.1142B} criteria \citep[with updates by][]{2015AJ....150..163B}: PSO~J$069.7303+04.3834$ (PSO~J$069+04$; T2), PSO~J$049.1159+26.8409$ (PSO~J$049+26$; T2.5), PSO~J$168.1800-27.2264$ (PSO~J$168-27$; T2.5), and ULAS~J$1316+0312$ (T3). We also find 6~``weak'' composite candidates, meeting $1-2$ criteria: 2MASS~J$2139+0220$ (T1.5), CFHT-Hy-20 (T2.5), SDSS~J$152103.24+013142.7$ (SDSS~J$1521+0131$; T3), 2MASS~J$00132229-1143006$ (2MASS~J$0013-1143$; T4), WISEPA~J$081958.05-033529.0$ (WISE~J$0819-0335$; T4), and WISE~J$163645.56-074325.1$ (WISE~J$1636-0743$; T4.5). Such spectral peculiarity for 4 out of these 10 candidates has also been noted by \cite{2010ApJ...710.1142B}, \cite{2015ApJ...814..118B}, and \cite{2017AJ....154..112K} using the same criteria. We note that all 10 of our composite candidates reside in the L/T transition, with spectral types of T1.5--T4.5.

\begin{figure*}[t]
\begin{center}
\includegraphics[height=1.6in]{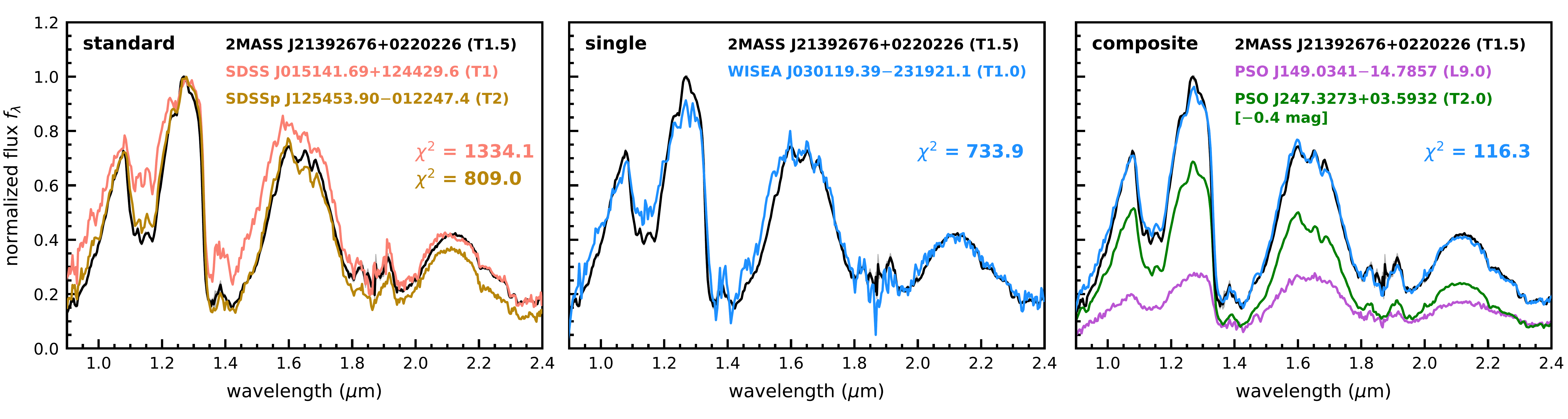}
\includegraphics[height=1.6in]{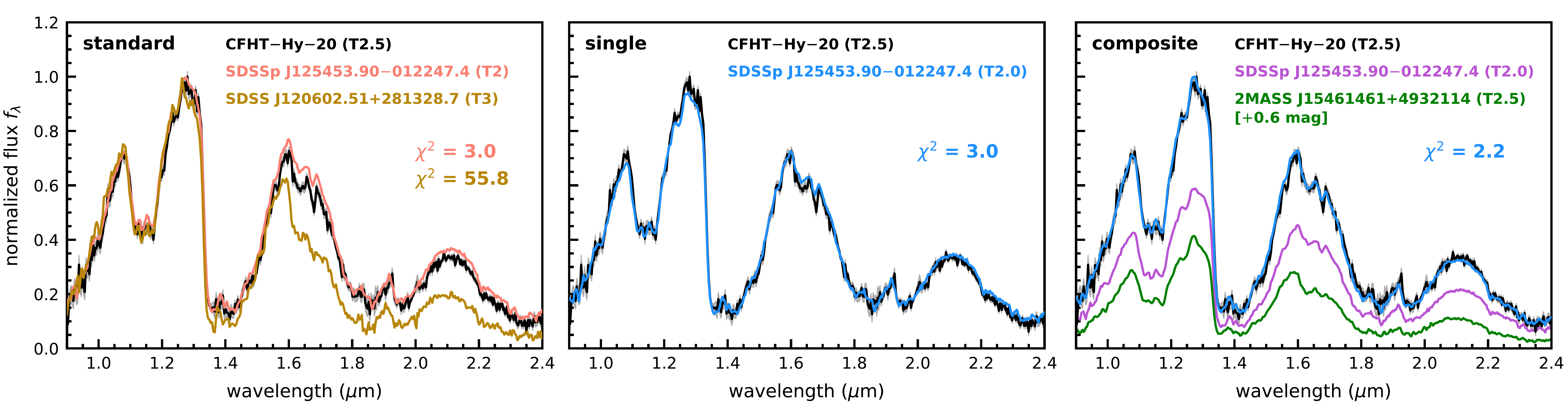}
\includegraphics[height=1.6in]{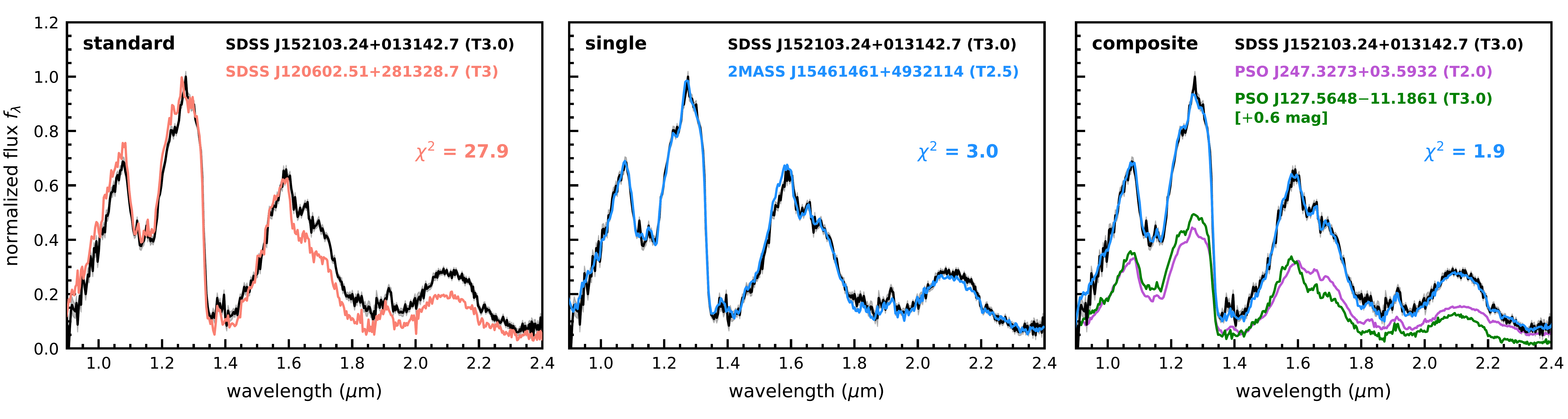}
\caption{Near-infrared spectra of our 6 weak composite candidates with the same format as Figure~\ref{fig:spec_strong}.}
\label{fig:spec_weak}
\end{center}
\end{figure*}
\renewcommand{\thefigure}{\arabic{figure}}
\addtocounter{figure}{-1}
\begin{figure*}[t]
\begin{center}
\includegraphics[height=1.6in]{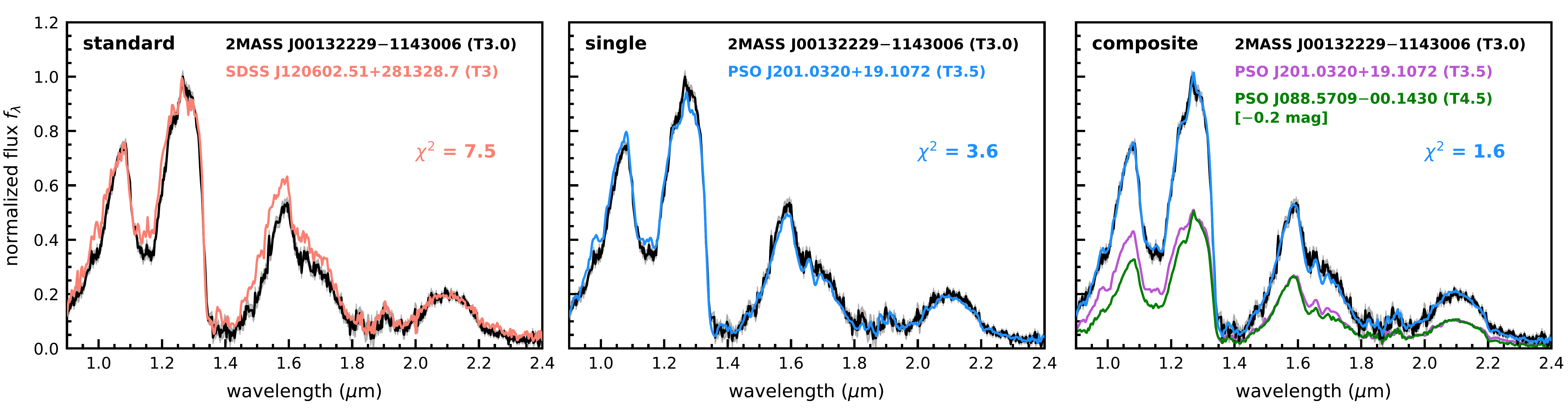}
\includegraphics[height=1.6in]{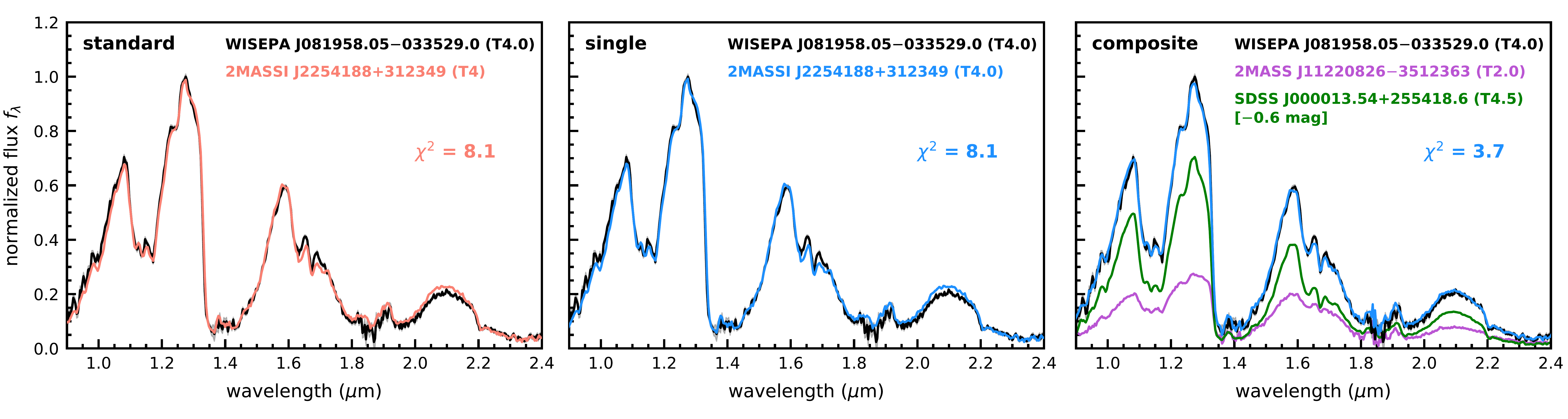}
\includegraphics[height=1.6in]{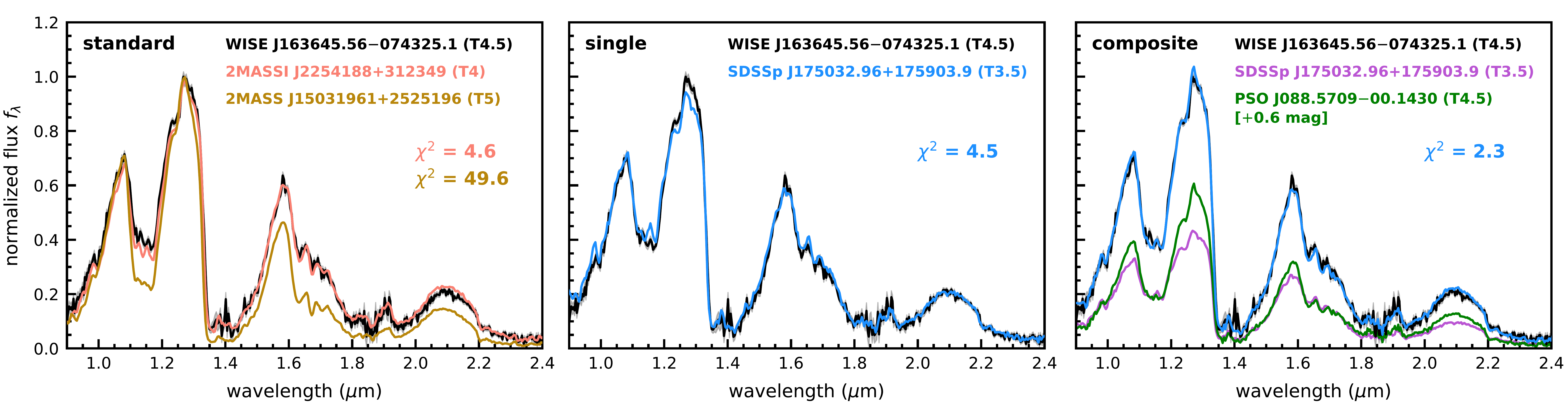}
\caption{Continued}
\label{fig:residual_afo_stgz}
\end{center}
\end{figure*}

We further perform spectral decomposition for our 10 candidates with peculiar spectra following the method described in \cite{2010ApJ...710.1142B}. We first construct empirical spectral templates by compiling all published IRTF/SpeX spectra of L5--T9 objects in The UltracoolSheet \citep[most of which are obtained from the SpeX Prism Library;][]{2014ASInC..11....7B}. We select objects that have median spectral S/N $\geqslant 30$ per pixel in $J$ band, as well as $K_{\rm MKO}$ magnitudes and parallaxes. We also exclude subdwarfs, resolved binaries, likely unresolved binaries identified in literature (using the same quantitative spectral indices as in this work), and all our identified YMG candidate members, leading to a total of 193 SpeX templates. We flux-calibrate each template using its $K_{\rm MKO}$-band absolute magnitude, with the WFCAM $K$-band filter and the corresponding zero-point flux from \cite{2006MNRAS.367..454H} and \cite{2007MNRAS.379.1599L}, respectively. This step is different from \cite{2010ApJ...710.1142B}, who estimated absolute magnitudes from spectral types using empirical relations. We then combine all possible pairs of flux-calibrated templates, resulting in 18,528 composite spectral templates. For the secondary component of each composite system, we further allow its absolute $K_{\rm MKO}$ magnitude to vary by 7 steps of $0$~mag, $\pm0.2$~mag, $\pm0.4$~mag, and $\pm0.6$~mag, use these magnitudes to flux-calibrate its spectrum again, and then generate 7 new composite spectral templates. These magnitude variations are chosen to account for the intrinsic $K_{\rm MKO}$-band photometric scatter of ultracool dwarfs at a given spectral types \citep[e.g., Figure~25 of][]{2012ApJS..201...19D}. We therefore obtain 129,696 composite spectral templates, expanded from the original set by a factor of 7. For a given binary candidate, we compare its spectrum with each composite template over wavelengths of $0.95 - 1.35$~$\mu$m, $1.45-1.8$~$\mu$m, and $2.0-2.35$~$\mu$m, and then compute the flux scale factor that minimizes the $\chi^{2}$ \citep[Equation~1 of][]{2010ApJ...710.1142B}. In addition to this synthetic composite fitting, we also fit the single-object spectral templates to our candidates' spectra with the same method. We have in total 246 single templates for such analysis as we do not require them to have parallaxes or $K_{\rm MKO}$ magnitudes. The best-fit single and composite templates for our 4 strong and 6 weak composite candidates are shown in Figures~\ref{fig:spec_strong} and \ref{fig:spec_weak}, respectively. 

We do not quantitatively assess whether composite templates provide better spectral matches than the single-object ones for our objects. Instead, we visually examine the best-fit single and composite templates to study the spectral peculiarity of each candidate. We find the best-fit single-object templates of 6~candidates do not match their observed spectra, 2MASS~J$0013-1143$, PSO~J$049+26$, PSO~J$069+04$, ULAS~J$1316+0312$, WISE~J$1636-0743$, and 2MASS~J$2139+0220$. These objects have (1) stronger CH$_{4}$ absorption at $1.6$~$\mu$m relative to $2.2$~$\mu$m (PSO~J$049+26$ and PSO~J$069+04$), (2) more prominent $J$-band peakS and/or deeper H$_{2}$O and CH$_{4}$ absorption around $1.1$~$\mu$m (2MASS~J$0013-1143$, PSO~J$069+04$, ULAS~J$1316+0312$, WISE~J$1636-0743$, 2MASS~J$2139+0220$), or (3) less prominent blue wingS of $Y$ band (2MASS~J$0013-1143$ and 2MASS~J$2139+0220$). The first two phenomena have also been seen in the integrated IRTF/SpeX spectra of binaries by \cite{2010ApJ...710.1142B}. These differences do not appear when examining the best-fit composite templates of these 6 candidates. For the remaining 4 composite candidates, we do not see significant improvements by switching from single-object to composite spectral fitting.

Besides the 20 candidate YMG members with IRTF/SpeX spectra, our remaining 10 candidates have spectra taken by a variety of other instruments \citep[][]{2006AJ....131.2722C, 2008MNRAS.390..304P, 2010MNRAS.406.1885B, 2011AJ....141..203A, 2012ApJ...753..156K, 2013MNRAS.433..457B, 2013MNRAS.430.1171D, 2013ApJS..205....6M}, including UKIRT/UIST, UKIRT/CGS4, VLT/X-shooter, Subaru/ICRS, Gemini/NIRI, and Keck/NIRSPEC. Most of these spectra have only partial wavelength coverage in the near-infrared (e.g., $JH$-band only) and spectroscopic follow-up is needed to obtain spectra with wider, contiguous wavelength coverage for detailed atmospheric analysis. Among these 10 objects, only ULAS~J$0047+1436$ has been flagged as a strong binary candidate, by \cite{2013MNRAS.430.1171D} based on VLT/X-Shooter spectra and the \cite{2010ApJ...710.1142B} spectral index criteria. \cite{2013MNRAS.430.1171D} also conducted spectral decomposition using templates from the SpeX Prism Library and found this object is well-fitted by a L8$+$T7 composite. 

Among our 30 candidate YMG members, photometric variability has been reported for two objects, 2MASS~J$0013-1143$ and 2MASS~J$2139+0220$. 2MASS~J$0013-1143$ has a peak-to-peak amplitude of $4.6\pm0.2\%$ in $J$ band \citep{2019A&A...629A.145E}. 2MASS~J$2139+0220$ is the most variable ultracool dwarf known to date, with a peak-to-peak amplitude of $26\pm1\%$ in $J$ band \citep[][]{2012ApJ...750..105R} and $11-12\%$ in {\it Spitzer}/IRAC [3.6] and [4.5] bands \citep[][]{2016ApJ...826....8Y, 2017ApJ...842...78V}. Therefore, the unusual spectral features of these two objects are likely related to their variability. Variability monitoring for all our remaining composite candidates would be helpful to further investigate these objects' spectral peculiarity. To summarize, we have identified a total of 7 T1.5--T4.5 candidates with peculiar spectra indicative of either atmospheric variability or unresolved binarity: 2MASS~J$0013-1143$, PSO~J$049+26$, PSO~J$069+04$, ULAS~J$1316+0312$, WISE~J$1636-0743$, 2MASS~J$2139+0220$, and ULAS~J$0047+1436$ (Table~\ref{tab:phys}).

\begin{figure*}[t]
\begin{center}
\includegraphics[height=6in]{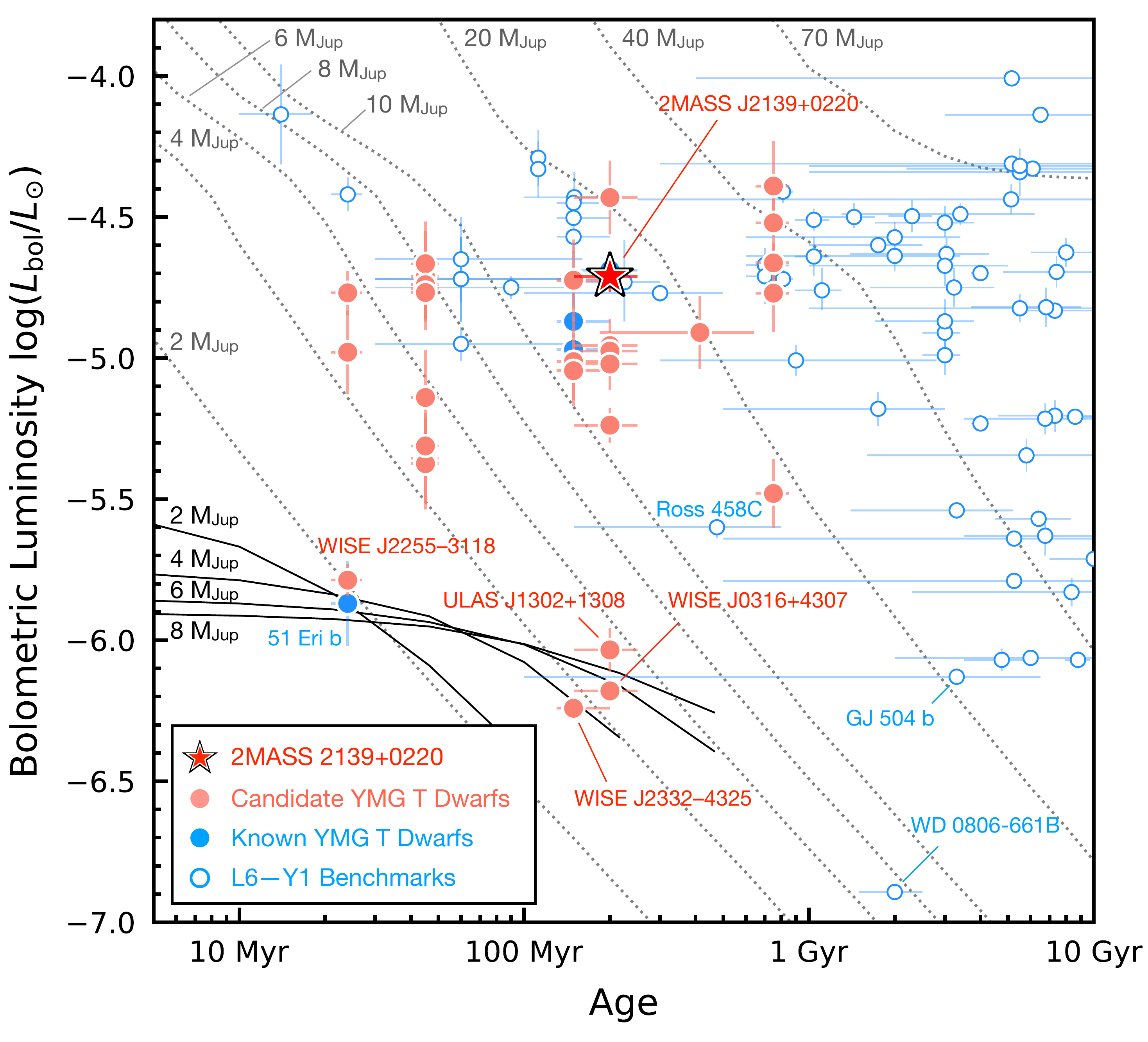}
\caption{Derived bolometric luminosities and ages of our identified candidates (light red), by assuming they are all YMG members. We use the red star to mark our newly confirmed Carina-Near member 2MASS~J$2139+0220$ and use blue solid circles for our recovered YMG T dwarfs. We overlay the L6$-$Y1 benchmarks (blue open circles) compiled by \cite{2020ApJ...891..171Z} and this work (Tables~\ref{tab:lateT_basic} and \ref{tab:lateT_phys}), the hot-start \cite{2008ApJ...689.1327S} hybrid evolutionary models (dashed grey lines), and the cold-start \cite{2008ApJ...683.1104F} evolutionary models (black solid lines). We find WISE~J$0316+4307$ (T8), WISE~J$2255-3118$ (T8), ULAS~J$1302+1308$ (T8.5), and WISE~J$2332-4325$ (T9) are all potential analogs of 51~Eri~b, having much fainter $L_{\rm bol}$ than the other benchmark ultracool dwarfs with similar ages. }
\label{fig:lbol_age}
\end{center}
\end{figure*}

\subsection{Physical Properties}
\label{subsec:physical}
We derive physical properties of our 30 candidates by assuming they are all YMG members. We first estimate the objects' bolometric luminosities from their broadband photometry. There are 19 objects with T0$-$T7 spectral types and parallaxes, and 12 of them have $K_{\rm 2MASS}$, which we convert into $L_{\rm bol}$ using the \cite{2015ApJ...810..158F} bolometric correction for young ultracool dwarfs based on the objects' spectral types. The other 7 ($=19 - 12$) objects do not have $K_{\rm 2MASS}$ data so we first convert their $K_{\rm MKO}$ (6 objects) or $J_{\rm MKO}$ (1 object) photometry into 2MASS photometry using their spectra and then apply the corresponding \cite{2015ApJ...810..158F} bolometric correction for young objects. We also have 4 T8$-$T9 candidates with parallaxes, and these objects' spectral types exceed the applicable range (M7$-$T7) of the \cite{2015ApJ...810..158F} bolometric corrections. Therefore, we use the super-magnitude method of \cite{2013Sci...341.1492D} as updated by W. Best et al. (in preparation). Briefly, this method computes bolometric luminosities for a set of absolute magnitudes composed of (1) $J_{\rm MKO}$, $H_{\rm MKO}$, {\it Spitzer}/IRAC $[3.6]$, and {\it Spitzer}/IRAC $[4.5]$ bands, (2) $J_{\rm MKO}$, $H_{\rm MKO}$, $W1$, $W2$ bands, or (3) subsets of the first two lists, using polynomials determined from the Sonora-Bobcat cloudless model atmospheres (\citealt{2017AAS...23031507M}; Marley et al. in prep). Among these 4 objects, WISE~J$2255-3118$ has IRTF/SpeX spectra and has been recently analyzed by us using Sonora-Bobcat models \citep[][]{Zhang2020c}. \cite{Zhang2020c} also computed this object's $L_{\rm bol}$ by integrating its observed $1.0-2.5$~$\mu$m SpeX spectrum and fitted model spectra to shorter and longer wavelengths spanning $0.4-50$~$\mu$m. The resulting spectroscopic $L_{\rm bol} = -5.72 \pm 0.03$~dex is consistent with our super-magnitude $L_{\rm bol} = -5.79 \pm 0.06$~dex adopted in this work. Given that the \cite{2015ApJ...810..158F} bolometric corrections were also based on integrating spectral energy distributions of objects, there does not seem to be significant systematics in $L_{\rm bol}$ between our subsets of T0$-$T7 and T8$-$T9 candidates. For the remaining 7 ($= 30 - 19 - 4$) objects without trigonometric parallaxes, we use their photometric distances and convert their $K_{\rm MKO}$-band absolute magnitudes into $K_{\rm 2MASS}$ band using their spectra or polynomials provided by \cite{2015ApJ...810..158F}. Then, we apply the \cite{2015ApJ...810..158F} bolometric correction for young ultracool dwarfs to compute these objects' bolometric luminosities. We have propagated all uncertainties in magnitudes, parallaxes, spectral types, and empirical relations into our resulting $L_{\rm bol}$ values in a Monte Carlo fashion. 

We adopt YMG ages of $149^{+51}_{-19}$~Myr for AB~Doradus \citep[][]{2015MNRAS.454..593B}, $40-50$~Myr for Argus \citep{2019ApJ...870...27Z}, $24 \pm 3$~Myr for $\beta$~Pictoris \citep[][]{2015MNRAS.454..593B}, $200 \pm 50$~Myr for Carina-Near \citep[][]{2006ApJ...649L.115Z}, $750 \pm 100$~Myr for Hyades \citep[][]{2015ApJ...807...24B}\footnote{The Hyades open cluster mentioned here is the core of, and thereby distinct from, the Hyades supercluster (a.k.a. Hyades stream or Hyades moving group) proposed by Olin Eggen \citep[e.g.,][]{1958MNRAS.118...65E}. It was initially hypothesized that members of the Hyades supercluster are coeval, but in fact this supercluster is composed of both young and field-age stars with a range of elemental abundances \citep[e.g.,][]{2001ASPC..228..398C, 2005A&A...430..165F, 2007A&A...461..957F, 2010ApJ...717..617B, 2011MNRAS.415..563D}. Throughout this work, we follow \cite{2018ApJ...856...23G} and use the Hyades designation to refer to the young open cluster \citep[$750 \pm 100$~Myr;][]{2015ApJ...807...24B}, with members compiled by \cite{1998A&A...331...81P}.}, and $414 \pm 23$~Myr for the Ursa Major cluster \citep[][]{2015ApJ...813...58J}. Combining our candidates' $L_{\rm bol}$ and ages, we then interpolate the hot-start \cite{2008ApJ...689.1327S} hybrid evolutionary models and compute their effective temperatures ($T_{\rm eff}$), surface gravities ($\log{g}$), radii ($R$), and masses ($M$) in a Monte Carlo fashion. We assume the objects' bolometric luminosities follow a normal distribution and assume their ages follow a uniform distribution (for Argus members) or a Gaussian distribution (for all other YMGs) constrained to $0-10$~Gyr. The derived physical properties of our 30 candidates are listed in Table~\ref{tab:phys} and shown in Figure~\ref{fig:lbol_age}. In total, 22 objects have planetary masses ($2-13$~M$_{\rm Jup}$) if their memberships are confirmed. 

We compare bolometric luminosities and ages of our candidates with all previously known L6$-$Y1 ultracool dwarfs with independently determined ages or masses (75 total benchmarks), including YMG members, wide-orbit companions to stars or white dwarfs, and ultracool binary systems with measured dynamical masses. We obtain properties of L6$-$T6 benchmarks (60 objects) from \cite{2020ApJ...891..171Z}, and we compile a catalog for T7$-$Y1 benchmarks (15 objects) in this work (Tables~\ref{tab:lateT_basic} and \ref{tab:lateT_phys}). We obtain $L_{\rm bol}$ of 11 T7$-$Y1 benchmarks from the literature, computed by integrating the objects' spectral energy distributions. Bolometric luminosities of the remaining 4 objects are either lacking (WISEU~J$005559.88+594745.0$, Wolf~1130C, WD~$0806-661$B) or from a model-based bolometric correction in $W2$ band \citep[WISE~J$111838.70+312537.9$; see][]{2013AJ....145...84W}, and therefore we (re-)compute their $L_{\rm bol}$ using the aforementioned super-magnitude method. 

Among these 15 T7$-$Y1 benchmarks, dynamical masses have been measured for Gl~229B \citep[][]{2020AJ....160..196B} and Gl~758B \citep[][]{2018AJ....155..159B, 2019AJ....158..140B}. Following the rejection sampling analysis in \cite{2017ApJS..231...15D}, \cite{2020AJ....160..196B} combined the dynamical mass and $L_{\rm bol}$ of Gl~229B to derive its age, $T_{\rm eff}$, $\log{g}$, and $R$ using the \cite{2008ApJ...689.1327S} hybrid evolutionary models. In this work, we conduct the same analysis for Gl~758B by using the more recent dynamical mass measured by \citeauthor{2019AJ....158..140B} (\citeyear{2019AJ....158..140B}; also see \citealt{2018AJ....155..159B, 2018A&A...615A.149C}). For the remaining 13 benchmarks with no independently inferred masses, we obtain their ages from their host stars as determined in the literature. We then derive these objects' $T_{\rm eff}$, $\log{g}$, $R$, and $M$ by using their $L_{\rm bol}$, ages, and the interpolated \cite{2008ApJ...689.1327S} hybrid evolutionary models as done for our YMG candidates. For the two benchmarks older than $\sim 10$~Gyr, WISEU~J$005559.88+594745.0$ \citep[$10\pm3$~Gyr;][]{2020ApJ...899..123M} and Wolf~1130C \citep[$>10$~Gyr;][]{2018ApJ...854..145M}, we derive their physical properties at an age of $10$~Gyr. 

As shown in Figure~\ref{fig:lbol_age}, we find our 4 latest-type candidates, WISE~J$0316+4307$ (T8), WISE~J$2255-3118$ (T8), ULAS~J$1302+1308$ (T8.5), and WISE~J$233226.49-432510.6$ (WISE~J$2332-4325$; T9) have comparably low luminosities and young ages as 51~Eri~b, with the bolometric luminosities being much fainter than other benchmark ultracool dwarfs with similar ages. We discuss these objects in the following section.

\begin{figure*}[t]
\begin{center}
\includegraphics[height=4.8in]{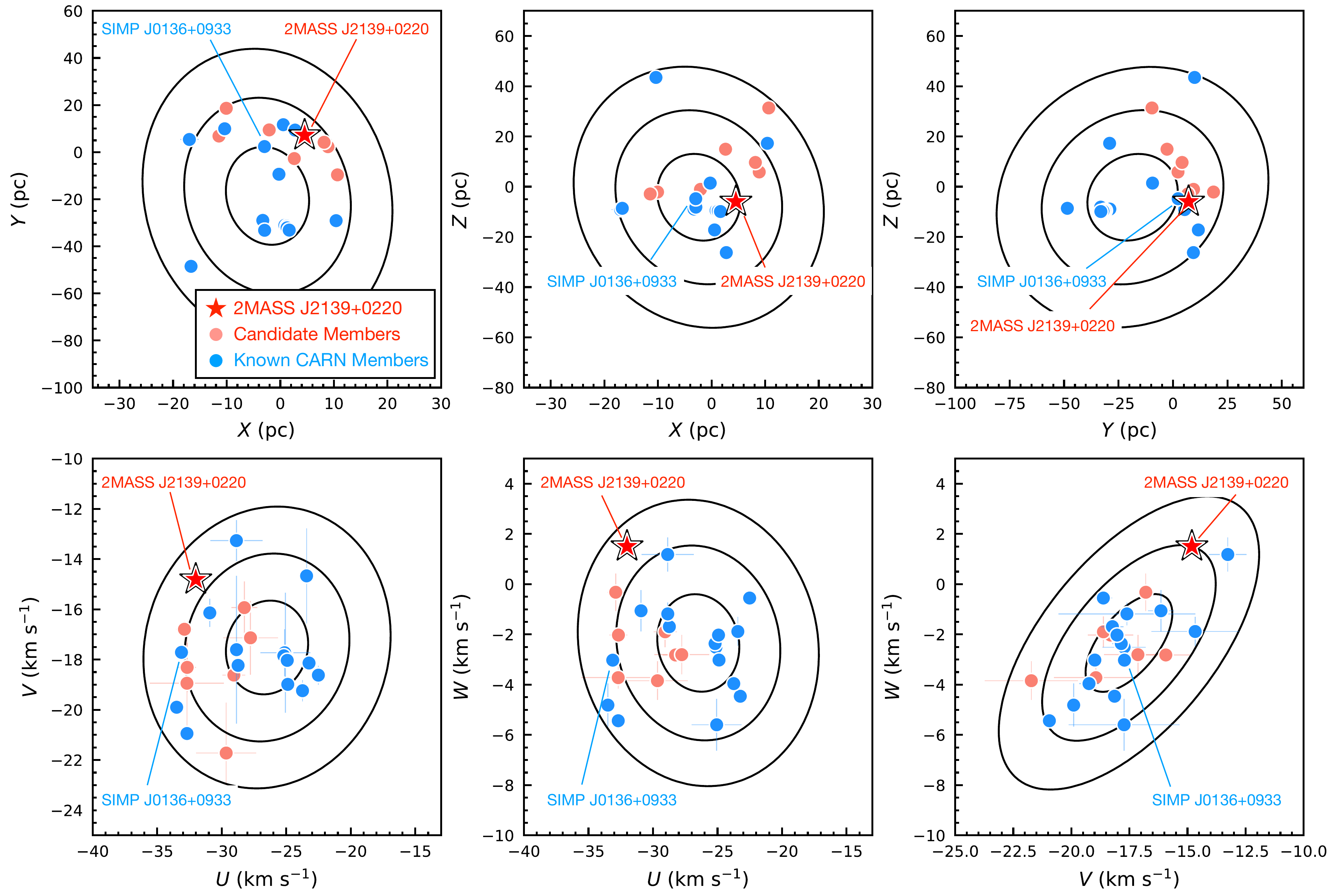}
\caption{Galactic $XYZUVW$ coordinates of our newly confirmed member 2MASS~J$2139+0220$ (red star), our new candidate members (light red circles), and previously known members (blue circles) of the Carina-Near moving group \citep[][]{2006ApJ...649L.115Z}. We obtain kinematic information of previously known members from \cite{2018ApJ...856...23G} and use the BANYAN~$\Sigma$ multivariate Gaussian model to generate $1\sigma/2\sigma/3\sigma$ extent of this group's $XYZUVW$ properties (black contours), with these contours encompassing $39.3\%$ ($1\sigma$), $86.5\%$ ($2\sigma$), and $98.9\%$ ($3\sigma$) of the cumulative volume for each bivariate Gaussian distribution. For our candidate members with no radial velocity, we use light red circles to mark their optimal $UVW$ space motions as Carina-Near members as computed by BANYAN~$\Sigma$. }
\label{fig:carn}
\end{center}
\end{figure*}

\subsection{Individual Notable Objects}
\label{subsec:discussion}

\subsubsection{2MASS~J$2139+0220$: A Newly Confirmed Member of the Carina-Near Moving Group}
\label{subsubsec:2M2139}
2MASS~J$2139+0220$ (T1.5) is the only object among our candidates that has full six-dimensional kinematic data (proper motion, parallax, and radial velocity). It has a high membership probability ($95.9\%$) in the Carina-Near moving group ($200 \pm 50$~Myr) based on BANYAN~$\Sigma$. Its membership probability is only $1\%$ based on LACEwING, which is known to produce a very low recovery rate for Carina-Near \citep[see Section~7.15 of][]{2017AJ....153...95R}. This object's $XYZUVW$ position lines up well with those of previously known Carina-Near members (Figure~\ref{fig:carn}), with $UVW$ space motion very close to a member of the Carina-Near stream (GJ~907.1; K8) as defined by \cite{2006ApJ...649L.115Z}.

Previous work has studied near-infrared spectra of 2MASS~J$2139+0220$ and derived its physical properties including surface gravity as an age indicator. \cite{2012ApJ...750..105R} fitted this object's IRTF/SpeX spectra (e.g., Figure~\ref{fig:spec_weak}) using the ultracool atmospheric models described in \cite{2008ApJ...678.1372C} and \cite{2009ApJ...702..154S} and developed by \cite{2001ApJ...556..872A}, \cite{2002ApJ...568..335M}, \cite{2008ApJ...689.1327S}, and they inferred $\log{g} = 4.5$~dex, which is the lowest $\log{g}$ value where the models are defined. \cite{2013ApJ...768..121A} compared this object's {\it HST}/WFC3 G141 grism spectra ($1.05 - 1.7$~$\mu$m, $R \sim 130$) with the \cite{2006ApJ...640.1063B} and the \cite{2011ASPC..448...91A} BT-Settl atmospheric models, and inferred $\log{g}$ of $4.0$~dex and $4.5$~dex, respectively. While these spectroscopically inferred surface gravities might contain modeling systematics, they are all consistent with our $\log{g} = 4.42^{+0.12}_{-0.06}$~dex based on evolutionary models and the assumption of its YMG membership (Section~\ref{subsec:physical}). More recent work by \cite{2017ApJ...842...78V} inferred a much higher $\log{g}$ of $5.37 \pm 0.02$~dex based on the Keck/NIRSPEC spectra ($2.29 - 2.33$~$\mu$m, $R \sim 25,000$), but such $\log{g}$ might be less reliable given the narrow wavelength coverage and the lack of gravity-sensitive lines in the data.

2MASS~J$2139+0220$ is the most variable ultracool dwarf known to date \citep[][]{2012ApJ...750..105R}, and the coexistence of its very high variability and potential young age (thereby low $\log{g}$) is in accord with a tentative correlation between low surface gravity and high-amplitude variability of mid-L and L/T transition dwarfs \citep[e.g.,][]{2015ApJ...799..154M, 2015ApJ...813L..23B, 2016ApJ...829L..32L, 2018MNRAS.474.1041V, 2018AJ....155..238S, 2020ApJ...893L..30B, 2020AJ....160...77Z}. Photometric variability has been also detected in previously known T-dwarf YMG members. SIMP~J013656.5+093347.3 (SIMP~J$0136+0933$) is a T2.5 member of the same Carina-Near moving group identified by \cite{2017ApJ...841L...1G} and has a $\approx 5\%$ variability in $J$ band \citep{2009ApJ...701.1534A, 2014ApJ...793...75R, 2017ApJ...842...78V}. GU~Psc~b \citep[][]{2014ApJ...787....5N} and 2MASS~J$1324+6358$ \citep[][]{2018ApJ...854L..27G} are both members of the AB~Doradus moving group, with variability of $\approx 4\%$ in $J$ band \citep[for GU~Psc~b detected by][]{2017AJ....154..129N} or $\approx 3\%$ in the mid-infrared \citep[for 2MASS~J$1324+6358$ detected by][]{2015ApJ...799..154M}. The high-amplitude variability of 2MASS~J$2139+0220$ is likely related to its peculiar spectrum (Figure~\ref{fig:spec_strong}) and might also have the impact on its redder, fainter near-infrared photometry compared to field dwarfs \citep[Figure~\ref{fig:cmd}; e.g.,][]{2020AJ....159..125L}. 

To summarize, we assign 2MASS~J$2139+0220$ as a new kinematic member of the Carina-Near moving group, making it the second T dwarf (after SIMP~J$0136+0933$ [T2.5]) in this group, as well as the third closest group member to Earth (after SIMP~J$0136+0933$ [$6.11 \pm 0.03$~pc] and GJ~358 [$9.601 \pm 0.004$~pc]).

\subsubsection{Young Late-T Candidates: WISE~J$0316+4307$, ULAS~J$1302+1308$, WISE~J$2255-3118$, and WISE~J$2332-4325$}
\label{subsubsec:fourlateT}
As seen in Figure~\ref{fig:lbol_age}, our 4 latest-type candidates have much fainter $L_{\rm bol}$ than other ultracool benchmarks with similar ages. WISE~J$0316+4307$ (T8) and ULAS~J$1302+1308$ (T8.5) are both candidate members of the Carina-Near moving group ($200 \pm 50$~Myr) with BANYAN~$\Sigma$ probabilities of $95.4\%$ and $97.7\%$, resepctively. Their membership probabilities are both $0\%$ based on LACEwING, which is known to produce a very low recovery rate for confirmed members of Carina-Near \citep[see Section~7.15 of][]{2017AJ....153...95R}. WISE~J$0316+4307$ has Keck/NIRSPEC spectra in $J$ and $H$ bands \citep[][]{2013ApJS..205....6M} which show no anomalies when compared to NIRSPEC spectra \citep[][]{2003ApJ...596..561M} of the T8 spectral standard, 2MASSI~J$0415195-093506$ \citep[2MASS~J$0415-0935$;][]{2006ApJ...637.1067B}. However, WISE~J$0316+4307$ has $\sim 2$~mag fainter $J$- and $H$-band absolute magnitudes than typical field T8 dwarfs. ULAS~J$1302+1308$ has much redder $J-K$ and $H-K$ colors than other T8$-$T9 field dwarfs (Figure~\ref{fig:cmd}), and its Subaru/IRCS spectrum has slightly enhanced fluxes near the $Y$-band peak as compared to 2MASS~J$0415-0935$, likely suggesting a lower surface gravity based on the Sonora-Bobcat cloudless models \citep[\citealt{2017AAS...23031507M}; Marley et al. in prep; e.g., see Figure~1 of][]{2020arXiv201112294Z}. 

WISE~J$2255-3118$ (T8) is a candidate member of the $\beta$~Pictoris moving group ($24 \pm 3$~Myr) with a $98.7\%$ BANYAN~$\Sigma$ probability and a $32\%$ LACEwING probability. Similar to ULAS~J$1302+1308$, WISE~J$2255-3118$ also has unusually red $J-K$ and $H-K$ colors (Figure~\ref{fig:cmd}), with a slightly enhanced $Y$-band peak flux compared to the T8 spectral standard 2MASS~J$0415-0935$, indicative of a lower surface gravity. Recently, we \citep{Zhang2020c} have analyzed the IRTF/SpeX spectra of late-T dwarfs using the cloudless Sonora-Bobcat models, with WISE~J$2255-3118$ included in our sample. Our spectroscopically inferred surface gravity $\log{g} = 3.66^{+0.32}_{-0.30}$~dex of WISE~J$2255-3118$ is among the lowest in our entire late-T dwarf sample, and is also consistent with its $\log{g} = 3.54^{+0.03}_{-0.02}$~dex in this work using the evolutionary models with assumed YMG membership (Section~\ref{subsec:physical} and Table~\ref{tab:phys}).

WISE~J$2332-4325$ (T9) is a candidate member of the AB~Doradus moving group ($149^{+51}_{-19}$~Myr) with a BANYAN~$\Sigma$ probability of $98.9\%$. This object has $J$-band Keck/NIRSPEC spectra \citep[][]{2018ApJS..236...28T} consistent with the T9 spectral standard UGPS~J$072227.51-054031.2$ \citep{2011ApJ...743...50C}. However, WISE~J$2332-4325$ has a much fainter $J$-band absolute magnitude and redder $J-H$ color than typical field T9 dwarfs (Figure~\ref{fig:cmd}).

The anomalous spectrophotometric appearance of these 4 objects is very similar to the $\beta$~Pictoris moving group exoplanet 51~Eri~b \citep[T6.5;][]{2015Sci...350...64M}, which also has unusually faint absolute magnitudes and red near-infrared colors.   Moreover, all these objects have distinctly faint $L_{\rm bol}$ as compared to other ultracool benchmarks with similar ages (Figure~\ref{fig:lbol_age}). Radial velocity follow-up is needed to assess the YMG membership of these 4 candidates, and if their young ages are confirmed, they will become the latest-type kinematic members of any young moving groups or associations. Most notably, they will be the first free-floating planets, whose physical properties are compatible with formation by both hot-start \citep[with high initial entropy and no subsequent accretion; e.g.,][]{1997ApJ...491..856B, 2001ApJ...554.1274C, 2003AandA...402..701B, 2008ApJ...689.1327S} and cold-start \citep[with low initial entropy and core accretion; e.g.,][]{2007ApJ...655..541M, 2008ApJ...683.1104F} conditions. These late-type YMG members may therefore shed insight on the formation pathways of directly-imaged and free-floating planets.

\section{Summary}
\label{sec:summary}
We have identified new and candidate T-dwarf members of nearby young moving using astrometry for 694 T and Y dwarfs, including 447 objects with parallaxes, mostly produced by recent large near-infrared astrometric programs by \cite{2019ApJS..240...19K} and \cite{2020AJ....159..257B}. Using the BANYAN~$\Sigma$ and LACEwING algorithms, we have recovered all 5 previously known T-dwarf YMG members and  identified 30 new candidate members.

We find 4 early-T (including 2MASS~J$2139+0220$) and 3 late-T candidate members exhibit $0.4-0.8$~mag redder $J-K$ colors and/or $0.6-2.2$~mag fainter $J$-band absolute magnitudes than field dwarfs with similar spectral types. Such anomalous photometry is in accord with previously known YMG T dwarfs (e.g., 2MASS~J$1324+6358$ and 51~Eri~b), providing evidence of their youth and thereby supporting YMG membership. 

Several of our candidates show unusual spectral features that differ from single ultracool dwarfs. Such peculiarities are consistent with unresolved binarity. Alternatively, these objects might have inhomogeneous cloud cover and/or temperature fluctuations, which cause their spectra to appear like the sum of photospheres with different effective temperatures. Variability monitoring would help to investigate their spectral peculiarities. 

We have estimated bolometric luminosities of all our candidates and inferred their effective temperatures, surface gravities, radii, and masses from evolutionary models assuming they are YMG members. The resulting mass estimates for 22 out of 30 candidates span $2-13$~M$_{\rm Jup}$, firmly in the planetary-mass regime. We establish the high-amplitude variable T1.5 dwarf 2MASS~J$2139+0220$ as a new planetary-mass member ($14.6^{+3.2}_{-1.6}$~M$_{\rm Jup}$) of the Carina-Near ($200 \pm 50$~Myr) moving group, making it the second T dwarf and the third closest member of this group. 2MASS~J$2139+0220$ is the most variable ultracool dwarf found to date, and the coexistence of its variability and youth is in accord with a tentative correlation between low surface gravity and high-amplitude variability of mid-L and L/T transition objects. Its high variability might also be related to its unusually red, faint near-infrared photometry and peculiar spectrum. 

Our four latest-type YMG candidates have spectral types of T8$-$T9. If confirmed, these objects will be the first free-floating planets whose ages and luminosities are compatible with both hot-start and cold-start evolutionary models, and thereby overlap the planetary-mass companion 51~Eri~b. The low surface gravity of these objects are also supported by their anomalous spectrophotometry and our recent atmospheric modeling for one of them (WISE~J$2255-3118$). Along with this analysis, we have also compiled all 15 previously known L7$-$Y1 benchmarks and derived a homogeneous set of their effective temperatures, surface gravities, radii, and masses.
 
Radial velocity measurements are needed to assess the membership of our YMG candidates except for 2MASS~J$2139+0220$. Such follow-up is feasible for our brightest candidates using existing high-resolution spectrographs on 8$-$10~meter-class telescopes (e.g., Gemini/GNIRS and Keck/NIRSPEC), but the majority of our candidates are too faint ($J \gtrsim 16.5$~mag, $K \gtrsim 15$~mag) and await 20$-$30~meter-class telescopes for radial velocity characterization.

\acknowledgments
This work has benefited from The UltracoolSheet at http://bit.ly/UltracoolSheet, maintained by Will Best, Trent Dupuy, Michael Liu, Rob Siverd, and Zhoujian Zhang, and developed from compilations by \cite{2012ApJS..201...19D}, \cite{2013Sci...341.1492D}, \cite{2016ApJ...833...96L}, \cite{2018ApJS..234....1B}, and \cite{2021AJ....161...42B}. M.C.L. acknowledges National Science Foundation (NSF) grant AST-1518339. This research was greatly facilitated by the TOPCAT software written by Mark Taylor (http://www.starlink.ac.uk/topcat/). Finally, the authors wish to recognize and acknowledge the very significant cultural role and reverence that the summit of Maunakea has always had within the indigenous Hawaiian community.  We are most fortunate to have the opportunity to conduct observations from this mountain.

\facilities{IRTF (SpeX), CFHT (WIRCam)}

\software{BANYAN~$\Sigma$ \citep[version~1.2; ][]{2018ApJ...856...23G}, LACEwING \citep[][]{2017AJ....153...95R}, TOPCAT \citep[][]{2005ASPC..347...29T}, Astropy \citep{2013A&A...558A..33A, 2018AJ....156..123A}, IPython \citep{PER-GRA:2007}, Numpy \citep{numpy},  Scipy \citep{scipy}, Matplotlib \citep{Hunter:2007}.}

\end{CJK*}

\clearpage
\bibliographystyle{aasjournal}
\bibliography{ms}

\clearpage
\tabletypesize{\scriptsize}

\global\pdfpageattr\expandafter{\the\pdfpageattr/Rotate 90} 
\begin{longrotatetable} 
\begin{deluxetable}{lccccccccccccccccc} 
\tablewidth{0pc} 
\setlength{\tabcolsep}{0.035in} 
\tablecaption{T-Dwarf Members and Candidates of Young Moving Groups \label{tab:members}} 
\tablehead{ \multicolumn{1}{l}{}  &  \multicolumn{1}{c}{}  &  \multicolumn{1}{c}{}  &  \multicolumn{1}{c}{}  &  \multicolumn{1}{c}{}  &  \multicolumn{1}{c}{}  &  \multicolumn{1}{c}{}  &  \multicolumn{1}{c}{}  &  \multicolumn{1}{c}{}  &  \multicolumn{2}{c}{Membership Probability}  &  \multicolumn{1}{c}{}  &  \multicolumn{6}{c}{References}  \\ 
\cline{10-11}  \cline{13-18} 
\multicolumn{1}{l}{Object}  &  \multicolumn{1}{c}{SpT}  &  \multicolumn{1}{c}{R.A.\tablenotemark{a}}  &  \multicolumn{1}{c}{Dec.\tablenotemark{a}}  &  \multicolumn{1}{c}{$\mu_{\alpha}\cos{\delta}$}  &  \multicolumn{1}{c}{$\mu_{\delta}$}  &  \multicolumn{1}{c}{Parallax\tablenotemark{b}}  &  \multicolumn{1}{c}{RV\tablenotemark{c}}  &  \multicolumn{1}{c}{}  &  \multicolumn{1}{c}{BANYAN $\Sigma$}  &  \multicolumn{1}{c}{LACEwING}  &  \multicolumn{1}{c}{}  &  \multicolumn{1}{c}{SpT}  &  \multicolumn{1}{c}{Coord.}  &  \multicolumn{1}{c}{PM}  &  \multicolumn{1}{c}{Parallax}  &  \multicolumn{1}{c}{RV}  &  \multicolumn{1}{c}{Membership}  \\ 
\multicolumn{1}{l}{}  &  \multicolumn{1}{c}{}  &  \multicolumn{1}{c}{(hh:mm:ss.ss)}  &  \multicolumn{1}{c}{(dd:mm:ss.ss)}  &  \multicolumn{1}{c}{(mas yr$^{-1}$)}  &  \multicolumn{1}{c}{(mas yr$^{-1}$)}  &  \multicolumn{1}{c}{(mas)}  &  \multicolumn{1}{c}{(km s$^{-1}$)}  &  \multicolumn{1}{c}{}  &  \multicolumn{1}{c}{}  &  \multicolumn{1}{c}{}  &  \multicolumn{1}{c}{}  &  \multicolumn{1}{c}{}  &  \multicolumn{1}{c}{}  &  \multicolumn{1}{c}{}  &  \multicolumn{1}{c}{}  &  \multicolumn{1}{c}{}  &  \multicolumn{1}{c}{}  } 
\startdata 
\cline{1-18} 
\multicolumn{18}{c}{AB Doradus} \\ 
\cline{1-18} 
\multicolumn{18}{l}{$\bullet$ New Candidate Members (with trigonometric parallax)} \\ 
 WISE J163645.56$-$074325.1  &  T4.5  &  16:36:45.65  &  $-$07:43:24.24  &  $-39.4 \pm 1.3$  &  $-152.4 \pm 1.7$  &  $33.0 \pm 5.0$  &  [$-16.7 \pm 1.4$]  &  &  $81.6\%$  &  $7.0\%$  &  &  21  &  31  &  36  &  41  &  $\cdots$  &  43  \\ 
 WISEPA J062720.07$-$111428.8  &  T6  &  06:27:20.09  &  $-$11:14:24.36  &  $-13.2 \pm 1.2$  &  $-337.8 \pm 1.1$  &  $75.0 \pm 4.0$  &  [$+25.0 \pm 1.0$]  &  &  $99.0\%$  &  $30.0\%$  &  &  13  &  31  &  40  &  40  &  $\cdots$  &  43  \\ 
 WISE J233226.49$-$432510.6\tablenotemark{d}  &  T9  &  23:32:26.54  &  $-$43:25:10.92  &  $249.7 \pm 1.2$  &  $-249.9 \pm 1.5$  &  $65.3 \pm 2.6$  &  [$+13.7 \pm 1.5$]  &  &  $98.9\%$  &  $8.0\%$  &  &  15  &  22  &  40  &  40  &  $\cdots$  &  43  \\ 
\multicolumn{18}{l}{$\bullet$ New Candidate Members (with photometric parallax)} \\ 
 ULAS J081918.58+210310.4  &  T6  &  08:19:18.62  &  21:03:12.60  &  $-58.0 \pm 11.0$  &  $-181.0 \pm 11.0$  &  [$33.2 \pm 3.2$]  &  [$+9.0 \pm 1.8$]  &  &  $86.3\%$  &  $12.0\%$  &  &  18  &  31  &  18  &  42  &  $\cdots$  &  43  \\ 
\multicolumn{18}{l}{$\bullet$ Recovered Previously-Known Members} \\ 
 2MASS J13243553+6358281\tablenotemark{d}  &  T2.5  &  13:24:35.50  &  63:58:27.84  &  $-368.0 \pm 4.0$  &  $-63.7 \pm 2.7$  &  $79.0 \pm 9.0$  &  $-23.7 \pm 0.4$  &  &  $99.8\%$  &  $64.0\%$  &  &  10  &  31  &  36  &  38  &  38  &  38  \\ 
 GU Psc b  &  T3.5  &  01:12:35.04  &  17:03:55.44  &  $96.64 \pm 0.13$  &  $-100.70 \pm 0.11$  &  $21.00 \pm 0.07$  &  $-1.5 \pm 0.5$  &  &  $99.1\%$  &  $72.0\%$  &  &  24  &  37  &  29,37  &  29,37  &  23  &  24  \\ 
 SDSSp J111010.01+011613.1  &  T5.5  &  11:10:10.01  &  01:16:12.72  &  $-217.1 \pm 0.7$  &  $-280.9 \pm 0.6$  &  $52.1 \pm 1.2$  &  $+7.5 \pm 3.8$  &  &  $99.3\%$  &  $46.0\%$  &  &  4  &  31  &  14  &  14  &  26  &  26  \\ 
\cline{1-18} 
\multicolumn{18}{c}{Argus} \\ 
\cline{1-18} 
\multicolumn{18}{l}{$\bullet$ New Candidate Members (with trigonometric parallax)} \\ 
 SDSS J152103.24+013142.7  &  T3  &  15:21:03.24  &  01:31:42.60  &  $-176.0 \pm 4.0$  &  $43.0 \pm 4.0$  &  $43.0 \pm 6.0$  &  [$-18.6 \pm 1.3$]  &  &  $82.3\%$  &  $0.0\%$  &  &  28  &  31  &  41  &  41  &  $\cdots$  &  43  \\ 
 2MASS J00132229$-$1143006  &  T4  &  00:13:22.32  &  $-$11:43:00.48  &  $214.0 \pm 3.0$  &  $-27.0 \pm 3.0$  &  $40.0 \pm 3.0$  &  [$+0.4 \pm 1.3$]  &  &  $96.8\%$  &  $0.0\%$  &  &  33  &  31  &  41  &  41  &  $\cdots$  &  43  \\ 
 SDSS J020742.48+000056.2  &  T4.5  &  02:07:42.84  &  00:00:55.80  &  $159.0 \pm 3.0$  &  $-14.0 \pm 4.0$  &  $29.0 \pm 4.0$  &  [$+12.6 \pm 1.3$]  &  &  $95.6\%$  &  $0.0\%$  &  &  4  &  31  &  11  &  11  &  $\cdots$  &  43  \\ 
 WISE J024124.73$-$365328.0\tablenotemark{d}  &  T7  &  02:41:24.74  &  $-$36:53:27.97  &  $242.0 \pm 1.5$  &  $148.4 \pm 1.4$  &  $52.4 \pm 2.7$  &  [$+11.8 \pm 1.5$]  &  &  $87.7\%$  &  $5.0\%$  &  &  39  &  22  &  40  &  40  &  $\cdots$  &  43  \\ 
\multicolumn{18}{l}{$\bullet$ New Candidate Members (with photometric parallax)} \\ 
 ULAS J004757.41+154641.4  &  T2  &  00:47:57.43  &  15:46:41.16  &  $147.0 \pm 28.0$  &  $-12.0 \pm 28.0$  &  [$27.0 \pm 2.9$]  &  [$+2.2 \pm 3.1$]  &  &  $80.1\%$  &  $0.0\%$  &  &  19  &  22  &  16  &  42  &  $\cdots$  &  43  \\ 
 PSO J168.1800$-$27.2264  &  T2.5  &  11:12:43.25  &  $-$27:13:36.12  &  $-120.0 \pm 120.0$  &  $100.0 \pm 40.0$  &  [$26.3 \pm 2.8$]  &  [$+8.2 \pm 6.7$]  &  &  $83.4\%$  &  $0.0\%$  &  &  25  &  22  &  25  &  42  &  $\cdots$  &  43  \\ 
 ULAS J154701.84+005320.3  &  T5.5  &  15:47:01.80  &  00:53:21.12  &  $-76.0 \pm 11.0$  &  $7.0 \pm 10.0$  &  [$23.3 \pm 2.7$]  &  [$-20.9 \pm 1.8$]  &  &  $95.3\%$  &  $0.0\%$  &  &  8  &  22  &  18  &  42  &  $\cdots$  &  43  \\ 
 ULAS J120744.65+133902.7  &  T6  &  12:07:44.62  &  13:39:02.88  &  $-155.0 \pm 12.0$  &  $1.0 \pm 12.0$  &  [$25.0 \pm 2.5$]  &  [$+1.5 \pm 1.5$]  &  &  $90.6\%$  &  $0.0\%$  &  &  9  &  22  &  18  &  42  &  $\cdots$  &  43  \\ 
 ULAS J075829.83+222526.7  &  T6.5  &  07:58:29.78  &  22:25:27.84  &  $-105.0 \pm 10.0$  &  $-57.0 \pm 11.0$  &  [$38.2 \pm 3.6$]  &  [$+22.4 \pm 1.4$]  &  &  $92.7\%$  &  $1.0\%$  &  &  18  &  22  &  18  &  42  &  $\cdots$  &  43  \\ 
\cline{1-18} 
\multicolumn{18}{c}{$\beta$ Pictoris} \\ 
\cline{1-18} 
\multicolumn{18}{l}{$\bullet$ New Candidate Members (with trigonometric parallax)} \\ 
 WISEPA J081958.05$-$033529.0  &  T4  &  08:19:58.18  &  $-$03:35:26.88  &  $-198.7 \pm 2.6$  &  $-166.5 \pm 2.2$  &  $72.0 \pm 3.0$  &  [$+16.1 \pm 1.0$]  &  &  $83.9\%$  &  $1.0\%$  &  &  13  &  31  &  41  &  41  &  $\cdots$  &  43  \\ 
 CFBDS J232304.41$-$015232.3  &  T6  &  23:23:04.39  &  $-$01:52:32.88  &  $93.3 \pm 1.5$  &  $-63.4 \pm 1.6$  &  $30.2 \pm 2.2$  &  [$-1.9 \pm 0.9$]  &  &  $89.1\%$  &  $1.0\%$  &  &  12  &  31  &  41  &  41  &  $\cdots$  &  43  \\ 
 WISEPC J225540.74$-$311841.8\tablenotemark{d}  &  T8  &  22:55:40.75  &  $-$31:18:42.12  &  $300.2 \pm 1.5$  &  $-162.1 \pm 2.2$  &  $71.0 \pm 4.0$  &  [$+1.6 \pm 0.9$]  &  &  $98.7\%$  &  $31.0\%$  &  &  13  &  22  &  40  &  40  &  $\cdots$  &  43  \\ 
\multicolumn{18}{l}{$\bullet$ Recovered Previously-Known Members} \\ 
 51 Eri b  &  T6.5  &  04:37:36.14  &  $-$02:28:24.60  &  $44.3 \pm 0.2$  &  $-63.8 \pm 0.2$  &  $33.58 \pm 0.14$  &  $+21.0 \pm 1.8$  &  &  $99.9\%$  &  $50.0\%$  &  &  34  &  29,37  &  29,37  &  29,37  &  6  &  1,27  \\ 
\cline{1-18} 
\multicolumn{18}{c}{Carina-Near} \\ 
\cline{1-18} 
\multicolumn{18}{l}{$\bullet$ {\bf Newly Confirmed Member}} \\ 
 {\bf 2MASS J21392676+0220226}  &  T1.5  &  21:39:27.10  &  02:20:24.00  &  $489.7 \pm 0.7$  &  $125.0 \pm 0.8$  &  $96.5 \pm 1.1$  &  $-25.1 \pm 0.3$  &  &  $95.9\%$  &  $1.0\%$  &  &  4  &  22  &  43  &  43  &  35  &  43  \\ 
\multicolumn{18}{l}{$\bullet$ New Candidate Members (with trigonometric parallax)} \\ 
 ULAS J131610.13+031205.5  &  T3  &  13:16:10.22  &  03:12:05.76  &  $-221.6 \pm 2.9$  &  $-20.0 \pm 3.0$  &  $29.0 \pm 2.4$  &  [$-6.6 \pm 0.8$]  &  &  $91.7\%$  &  $0.0\%$  &  &  28  &  31  &  41  &  41  &  $\cdots$  &  43  \\ 
 PSO J004.6359+56.8370  &  T4.5  &  00:18:32.57  &  56:50:12.84  &  $376.0 \pm 3.0$  &  $10.4 \pm 2.7$  &  $47.0 \pm 4.0$  &  [$-0.7 \pm 1.0$]  &  &  $90.8\%$  &  $0.0\%$  &  &  41  &  31  &  41  &  41  &  $\cdots$  &  43  \\ 
 WISE J223617.59+510551.9\tablenotemark{d}  &  T5  &  22:36:16.80  &  51:05:48.12  &  $708.6 \pm 2.1$  &  $326.7 \pm 2.5$  &  $102.8 \pm 1.9$  &  [$-10.6 \pm 1.0$]  &  &  $98.0\%$  &  $0.0\%$  &  &  17  &  31  &  41  &  41  &  $\cdots$  &  43  \\ 
 SDSSp J162414.37+002915.6  &  T6  &  16:24:14.33  &  00:29:15.72  &  $-372.9 \pm 1.6$  &  $-9.1 \pm 2.0$  &  $91.8 \pm 1.2$  &  [$-28.7 \pm 1.1$]  &  &  $99.0\%$  &  $0.0\%$  &  &  4  &  31  &  2  &  2,40  &  $\cdots$  &  43  \\ 
 2MASSI J1553022+153236  &  T7  &  15:53:02.21  &  15:32:36.96  &  $-385.9 \pm 0.7$  &  $166.2 \pm 0.9$  &  $75.1 \pm 0.9$  &  [$-25.7 \pm 1.0$]  &  &  $89.6\%$  &  $0.0\%$  &  &  4  &  31  &  14  &  14  &  $\cdots$  &  43  \\ 
 WISE J031624.35+430709.1\tablenotemark{d}  &  T8  &  03:16:24.41  &  43:07:08.76  &  $372.4 \pm 1.5$  &  $-225.9 \pm 1.5$  &  $73.3 \pm 2.8$  &  [$+16.3 \pm 1.0$]  &  &  $95.4\%$  &  $0.0\%$  &  &  21  &  22  &  40  &  40  &  $\cdots$  &  43  \\ 
 ULAS J130217.21+130851.2  &  T8.5  &  13:02:17.09  &  13:08:51.00  &  $-445.0 \pm 6.0$  &  $5.0 \pm 7.0$  &  $65.0 \pm 5.0$  &  [$-4.4 \pm 0.8$]  &  &  $97.7\%$  &  $0.0\%$  &  &  9  &  22  &  20  &  20  &  $\cdots$  &  43  \\ 
\multicolumn{18}{l}{$\bullet$ Recovered Previously-Known Members} \\ 
 SIMP J013656.5+093347.3  &  T2.5  &  01:36:56.57  &  09:33:47.16  &  $1239.0 \pm 1.2$  &  $-17.4 \pm 0.8$  &  $163.7 \pm 0.7$  &  $+11.5 \pm 0.4$  &  &  $95.6\%$  &  $0.0\%$  &  &  3  &  29,37  &  29,37  &  29,37  &  32  &  32  \\ 
\cline{1-18} 
\multicolumn{18}{c}{Hyades} \\ 
\cline{1-18} 
\multicolumn{18}{l}{$\bullet$ New Candidate Members (with trigonometric parallax)} \\ 
 PSO J069.7303+04.3834  &  T2  &  04:38:55.18  &  04:23:00.24  &  $132.0 \pm 4.0$  &  $10.0 \pm 3.0$  &  $37.0 \pm 6.0$  &  [$+39.4 \pm 1.8$]  &  &  $84.7\%$  &  $78.0\%$  &  &  41  &  31  &  41  &  41  &  $\cdots$  &  43  \\ 
 PSO J049.1159+26.8409  &  T2.5  &  03:16:27.60  &  26:50:27.96  &  $201.1 \pm 2.4$  &  $-52.8 \pm 1.9$  &  $34.0 \pm 3.0$  &  [$+29.5 \pm 1.8$]  &  &  $80.3\%$  &  $45.0\%$  &  &  25  &  31  &  41  &  41  &  $\cdots$  &  43  \\ 
 PSO J052.2746+13.3754  &  T3.5  &  03:29:05.66  &  13:22:31.80  &  $273.0 \pm 2.0$  &  $-20.7 \pm 2.0$  &  $44.0 \pm 3.0$  &  [$+32.5 \pm 1.8$]  &  &  $92.8\%$  &  $63.0\%$  &  &  41  &  31  &  41  &  41  &  $\cdots$  &  43  \\ 
\multicolumn{18}{l}{$\bullet$ New Candidate Members (with photometric parallax)} \\ 
 WISEPA J030724.57+290447.6  &  T6.5  &  03:07:24.60  &  29:04:47.64  &  $100.0 \pm 300.0$  &  $-100.0 \pm 300.0$  &  [$38.5 \pm 3.7$]  &  [$+27.3 \pm 0.2$]  &  &  $39.3\%$  &  $82.0\%$\tablenotemark{e}  &  &  13  &  22  &  13  &  42  &  $\cdots$  &  43  \\ 
\multicolumn{18}{l}{$\bullet$ Recovered Previously-Known Candidate Members} \\ 
 CFHT$-$Hy$-$20  &  T2.5  &  04:30:38.71  &  13:09:56.88  &  $142.6 \pm 1.6$  &  $-16.5 \pm 1.7$  &  $30.8 \pm 1.5$  &  [$+40.4 \pm 2.0$]  &  &  $98.7\%$  &  $99.0\%$  &  &  30  &  31  &  30  &  30  &  $\cdots$  &  7  \\ 
\cline{1-18} 
\multicolumn{18}{c}{Ursa Major} \\ 
\cline{1-18} 
\multicolumn{18}{l}{$\bullet$ New Candidate Members (with trigonometric parallax)} \\ 
 SDSS J125011.65+392553.9  &  T4  &  12:50:11.71  &  39:25:55.55  &  $-42.0 \pm 3.0$  &  $-830.5 \pm 2.6$  &  $43.0 \pm 3.0$  &  [$-10.1 \pm 2.2$]  &  &  $0.0\%$  &  $78.0\%$  &  &  5  &  31  &  41  &  41  &  $\cdots$  &  43  \\ 
\enddata 
\tablenotetext{a}{Coordinates are provided at epoch J2000 with equinox J2000.} 
\tablenotetext{b}{Parallaxes inside brackets are derived from photometric distances.} 
\tablenotetext{c}{Radial velocities inside brackets are optimal values with the assumed YMG membership as inferred by BANYAN~$\Sigma$ (all candidates with membership probabilities $>80\%$) or LACEwING (only for WISEPA~J$030724.57+290447.6$ and SDSS~J$125011.65+392553.9$).} 
\tablenotetext{d}{These six objects (5 YMG candidates and 1 previously known YMG member) also have parallaxes and proper motions from \cite{2020arXiv201111616K}, which became available while our paper was under review. The new astrometry does not alter the candidacy of these objects, but does slightly change their BANYAN $\Sigma$ membership probabilities to $99.4\%$ (AB Doradus) for 2MASS~J13243553$+$6358281, $95.5\%$ (Argus) for WISE~J024124.73$-$365328.0, $94.4\%$ (Carina-Near) for WISE~J031624.35$+$430709.1, $90.5\%$ (Carina-Near) for WISE~J223617.59$+$510551.9, $99.7\%$ (AB Doradus) for WISE~J233226.49$-$432510.6, and $99.1\%$ ($\beta$~Pictoris) for WISEPC~J225540.74$-$311841.8. Also, 3 objects have their LACEwING membership probabilities changed to $55\%$ (AB Doradus) for WISE~J233226.49$-$432510.6, $62\%$ (AB Doradus) for 2MASS~J13243553$+$6358281, and $31\%$ ($\beta$~Pictoris) for WISEPC~J225540.74$-$311841.8.  } 
\tablenotetext{e}{Based on LACEwING, this object is also likely a member of the AB Doradus moving group with a probability of $40\%$.} 
\tablerefs{(1) \cite{2001ApJ...562L..87Z}, (2) \cite{2003AJ....126..975T}, (3) \cite{2006ApJ...651L..57A}, (4) \cite{2006ApJ...637.1067B}, (5) \cite{2006AJ....131.2722C}, (6) \cite{2007AN....328..889K}, (7) \cite{2008AandA...481..661B}, (8) \cite{2008MNRAS.390..304P}, (9) \cite{2010MNRAS.406.1885B}, (10) \cite{2010ApJS..190..100K}, (11) \cite{2010AandA...524A..38M}, (12) \cite{2011AJ....141..203A}, (13) \cite{2011ApJS..197...19K}, (14) \cite{2012ApJS..201...19D}, (15) \cite{2012ApJ...753..156K}, (16) \cite{2012yCat.2314....0L}, (17) \cite{2013ApJ...777...84B}, (18) \cite{2013MNRAS.433..457B}, (19) \cite{2013MNRAS.430.1171D}, (20) \cite{2013AandA...560A..52M}, (21) \cite{2013ApJS..205....6M}, (22) \cite{2014yCat.2328....0C}, (23) \cite{2014ApJ...788...81M}, (24) \cite{2014ApJ...787....5N}, (25) \cite{2015ApJ...814..118B}, (26) \cite{2015ApJ...808L..20G}, (27) \cite{2015Sci...350...64M}, (28) \cite{2015MNRAS.449.3651M}, (29) \cite{2016AandA...595A...1G}, (30) \cite{2016ApJ...833...96L}, (31) \cite{2016arXiv161205242M}, (32) \cite{2017ApJ...841L...1G}, (33) \cite{2017AJ....154..112K}, (34) \cite{2017AJ....154...10R}, (35) \cite{2017ApJ...842...78V}, (36) \cite{2018ApJS..234....1B}, (37) \cite{2018AandA...616A...1G}, (38) \cite{2018ApJ...854L..27G}, (39) \cite{2018ApJS..236...28T}, (40) \cite{2019ApJS..240...19K}, (41) \cite{2020AJ....159..257B}, (42) \cite{ultracoolsheet}, (43) This Work} 
\end{deluxetable} 
\end{longrotatetable}

\global\pdfpageattr\expandafter{\the\pdfpageattr/Rotate 90} 
\begin{longrotatetable} 
\begin{deluxetable}{lccccccccccccccc} 
\tablewidth{0pc} 
\setlength{\tabcolsep}{0.05in} 
\tablecaption{Photometry of T-Dwarf YMG Members and Candidates \label{tab:phot}} 
\tablehead{ \multicolumn{1}{l}{}  &  \multicolumn{1}{c}{}  &  \multicolumn{1}{c}{}  &  \multicolumn{5}{c}{Near-Infrared MKO Photometry}  &  \multicolumn{1}{c}{}  &  \multicolumn{3}{c}{AllWISE Photometry}  &  \multicolumn{1}{c}{}  &  \multicolumn{3}{c}{{\it Spitzer}/IRAC Photometry}  \\ 
\cline{4-8}  \cline{10-12}  \cline{14-16}  
\multicolumn{1}{l}{Object}  &  \multicolumn{1}{c}{SpT}  &  \multicolumn{1}{c}{}  &  \multicolumn{1}{c}{$Y_{\rm MKO}$}  &  \multicolumn{1}{c}{$J_{\rm MKO}$}  &  \multicolumn{1}{c}{$H_{\rm MKO}$}  &  \multicolumn{1}{c}{$K_{\rm MKO}$}  &  \multicolumn{1}{c}{References}  &  \multicolumn{1}{c}{}  &  \multicolumn{1}{c}{$W1$}  &  \multicolumn{1}{c}{$W2$}  &  \multicolumn{1}{c}{References}  &  \multicolumn{1}{c}{}  &  \multicolumn{1}{c}{[3.6]}  &  \multicolumn{1}{c}{[4.5]}  &  \multicolumn{1}{c}{References}  \\ 
\multicolumn{1}{l}{}  &  \multicolumn{1}{c}{}  &  \multicolumn{1}{c}{}  &  \multicolumn{1}{c}{(mag)}  &  \multicolumn{1}{c}{(mag)}  &  \multicolumn{1}{c}{(mag)}  &  \multicolumn{1}{c}{(mag)}  &  \multicolumn{1}{c}{}  &  \multicolumn{1}{c}{}  &  \multicolumn{1}{c}{(mag)}  &  \multicolumn{1}{c}{(mag)}  &  \multicolumn{1}{c}{}  &  \multicolumn{1}{c}{}  &  \multicolumn{1}{c}{(mag)}  &  \multicolumn{1}{c}{(mag)}  &  \multicolumn{1}{c}{}  } 
\startdata 
\cline{1-16} 
\multicolumn{16}{c}{AB Doradus} \\ 
\cline{1-16} 
\multicolumn{16}{l}{$\bullet$ New Candidate Members (with trigonometric parallax)} \\ 
 WISE J163645.56$-$074325.1  &  T4.5  &  &  $17.60 \pm 0.05$  &  $16.42 \pm 0.02$  &  $16.28 \pm 0.05$  &  $16.32 \pm 0.05$  &  25,26  &  &  $15.93 \pm 0.06$  &  $14.68 \pm 0.06$  &  16  &  &  --  &  --  &  $\cdots$  \\ 
 WISEPA J062720.07$-$111428.8  &  T6  &  &  $16.37 \pm 0.07$  &  $15.25 \pm 0.05$  &  $15.50 \pm 0.18$  &  $15.51 \pm 0.18$  &  26  &  &  $14.98 \pm 0.03$  &  $13.25 \pm 0.03$  &  16  &  &  $14.27 \pm 0.02$  &  $13.32 \pm 0.02$  &  9  \\ 
 WISE J233226.49$-$432510.6  &  T9  &  &  --  &  $19.40 \pm 0.10$  &  $19.40 \pm 0.18$  &  --  &  10,18  &  &  $17.97 \pm 0.24$  &  $14.96 \pm 0.07$  &  16  &  &  $17.27 \pm 0.06$  &  $15.01 \pm 0.02$  &  24  \\ 
\multicolumn{16}{l}{$\bullet$ New Candidate Members (with photometric parallax)} \\ 
 ULAS J081918.58+210310.4  &  T6  &  &  $18.25 \pm 0.03$  &  $16.954 \pm 0.011$  &  $17.28 \pm 0.04$  &  $17.18 \pm 0.06$  &  11  &  &  $16.95 \pm 0.11$  &  $15.24 \pm 0.09$  &  16  &  &  --  &  --  &  $\cdots$  \\ 
\multicolumn{16}{l}{$\bullet$ Recovered Previously-Known Members} \\ 
 2MASS J13243553+6358281  &  T2.5  &  &  $16.52 \pm 0.08$  &  $15.44 \pm 0.07$  &  $14.68 \pm 0.06$  &  $14.08 \pm 0.06$  &  26  &  &  $13.12 \pm 0.02$  &  $12.29 \pm 0.02$  &  16  &  &  $12.56 \pm 0.03$  &  $12.33 \pm 0.03$  &  5  \\ 
 GU Psc b  &  T3.5  &  &  $19.40 \pm 0.05$  &  $18.12 \pm 0.03$  &  $17.70 \pm 0.03$  &  $17.40 \pm 0.03$  &  17  &  &  $17.17 \pm 0.33$  &  $15.41 \pm 0.22$  &  17  &  &  --  &  --  &  $\cdots$  \\ 
 SDSSp J111010.01+011613.1  &  T5.5  &  &  $17.338 \pm 0.012$  &  $16.161 \pm 0.008$  &  $16.20 \pm 0.02$  &  $16.05 \pm 0.03$  &  11  &  &  $15.44 \pm 0.04$  &  $13.92 \pm 0.04$  &  16  &  &  --  &  --  &  $\cdots$  \\ 
\cline{1-16} 
\multicolumn{16}{c}{Argus} \\ 
\cline{1-16} 
\multicolumn{16}{l}{$\bullet$ New Candidate Members (with trigonometric parallax)} \\ 
 SDSS J152103.24+013142.7  &  T3  &  &  $17.34 \pm 0.02$  &  $16.097 \pm 0.010$  &  $15.679 \pm 0.009$  &  $15.568 \pm 0.015$  &  11  &  &  $14.90 \pm 0.03$  &  $13.94 \pm 0.04$  &  16  &  &  --  &  --  &  $\cdots$  \\ 
 2MASS J00132229$-$1143006  &  T4  &  &  --  &  $16.05 \pm 0.02$  &  $15.74 \pm 0.22$  &  $15.76 \pm 0.22$  &  25,26  &  &  $15.49 \pm 0.05$  &  $14.32 \pm 0.05$  &  16  &  &  --  &  --  &  $\cdots$  \\ 
 SDSS J020742.48+000056.2  &  T4.5  &  &  $18.02 \pm 0.03$  &  $16.730 \pm 0.013$  &  $16.81 \pm 0.04$  &  $16.72 \pm 0.05$  &  11  &  &  $16.30 \pm 0.06$  &  $15.05 \pm 0.07$  &  16  &  &  --  &  --  &  $\cdots$  \\ 
 WISE J024124.73$-$365328.0  &  T7  &  &  --  &  $16.59 \pm 0.04$  &  $17.04 \pm 0.07$  &  --  &  23,24  &  &  $16.86 \pm 0.08$  &  $14.35 \pm 0.04$  &  16  &  &  $15.74 \pm 0.03$  &  $14.35 \pm 0.02$  &  24  \\ 
\multicolumn{16}{l}{$\bullet$ New Candidate Members (with photometric parallax)} \\ 
 ULAS J004757.41+154641.4  &  T2  &  &  $19.12 \pm 0.07$  &  $17.83 \pm 0.05$  &  $17.16 \pm 0.05$  &  $16.42 \pm 0.04$  &  11  &  &  $15.52 \pm 0.04$  &  $14.86 \pm 0.07$  &  16  &  &  --  &  --  &  $\cdots$  \\ 
 PSO J168.1800$-$27.2264  &  T2.5  &  &  $18.38 \pm 0.04$  &  $17.12 \pm 0.03$  &  $16.75 \pm 0.03$  &  $16.65 \pm 0.06$  &  15,26  &  &  $15.72 \pm 0.05$  &  $14.98 \pm 0.07$  &  16  &  &  --  &  --  &  $\cdots$  \\ 
 ULAS J154701.84+005320.3  &  T5.5  &  &  $19.37 \pm 0.06$  &  $18.32 \pm 0.03$  &  $18.45 \pm 0.07$  &  $18.21 \pm 0.10$  &  6  &  &  $16.88 \pm 0.10$  &  $15.89 \pm 0.15$  &  16  &  &  --  &  --  &  $\cdots$  \\ 
 ULAS J120744.65+133902.7  &  T6  &  &  $19.19 \pm 0.05$  &  $18.28 \pm 0.05$  &  $18.52 \pm 0.05$  &  $18.67 \pm 0.05$  &  7  &  &  $17.90 \pm 0.25$  &  $15.88 \pm 0.14$  &  16  &  &  --  &  --  &  $\cdots$  \\ 
 ULAS J075829.83+222526.7  &  T6.5  &  &  $18.68 \pm 0.04$  &  $17.62 \pm 0.02$  &  $17.91 \pm 0.02$  &  $17.87 \pm 0.12$  &  13  &  &  $16.52 \pm 0.09$  &  $15.07 \pm 0.08$  &  16  &  &  --  &  --  &  $\cdots$  \\ 
\cline{1-16} 
\multicolumn{16}{c}{$\beta$ Pictoris} \\ 
\cline{1-16} 
\multicolumn{16}{l}{$\bullet$ New Candidate Members (with trigonometric parallax)} \\ 
 WISEPA J081958.05$-$033529.0  &  T4  &  &  $15.94 \pm 0.05$  &  $14.78 \pm 0.02$  &  $14.60 \pm 0.05$  &  $14.64 \pm 0.05$  &  25,26  &  &  $14.35 \pm 0.03$  &  $13.08 \pm 0.03$  &  16  &  &  $13.61 \pm 0.02$  &  $13.07 \pm 0.02$  &  9  \\ 
 CFBDS J232304.41$-$015232.3  &  T6  &  &  $18.30 \pm 0.02$  &  $17.23 \pm 0.03$  &  $17.46 \pm 0.04$  &  $17.30 \pm 0.03$  &  8  &  &  $16.62 \pm 0.09$  &  $15.07 \pm 0.09$  &  16  &  &  --  &  --  &  $\cdots$  \\ 
 WISEPC J225540.74$-$311841.8  &  T8  &  &  $18.38 \pm 0.02$  &  $17.334 \pm 0.011$  &  $17.66 \pm 0.03$  &  $17.42 \pm 0.05$  &  20,26  &  &  $16.55 \pm 0.08$  &  $14.16 \pm 0.05$  &  16  &  &  $15.91 \pm 0.03$  &  $14.21 \pm 0.02$  &  9  \\ 
\multicolumn{16}{l}{$\bullet$ Recovered Previously-Known Members} \\ 
 51 Eri b  &  T6.5  &  &  --  &  $19.04 \pm 0.40$  &  $18.99 \pm 0.21$  &  $18.67 \pm 0.19$  &  22  &  &  --  &  --  &  $\cdots$  &  &  --  &  --  &  $\cdots$  \\ 
\cline{1-16} 
\multicolumn{16}{c}{Carina-Near} \\ 
\cline{1-16} 
\multicolumn{16}{l}{$\bullet$ {\bf Newly Confirmed Member}} \\ 
 {\bf 2MASS J21392676+0220226}  &  T1.5  &  &  $16.23 \pm 0.07$  &  $15.10 \pm 0.05$  &  $14.27 \pm 0.05$  &  $13.60 \pm 0.05$  &  26  &  &  $12.76 \pm 0.02$  &  $12.00 \pm 0.02$  &  16  &  &  --  &  --  &  $\cdots$  \\ 
\multicolumn{16}{l}{$\bullet$ New Candidate Members (with trigonometric parallax)} \\ 
 ULAS J131610.13+031205.5  &  T3  &  &  $18.00 \pm 0.03$  &  $16.75 \pm 0.02$  &  $16.13 \pm 0.02$  &  $15.43 \pm 0.02$  &  11  &  &  $14.16 \pm 0.03$  &  $13.76 \pm 0.04$  &  16  &  &  --  &  --  &  $\cdots$  \\ 
 PSO J004.6359+56.8370  &  T4.5  &  &  --  &  $16.22 \pm 0.02$  &  $16.24 \pm 0.02$  &  $16.13 \pm 0.02$  &  25  &  &  --  &  --  &  $\cdots$  &  &  --  &  --  &  $\cdots$  \\ 
 WISE J223617.59+510551.9  &  T5  &  &  $15.655 \pm 0.014$  &  $14.457 \pm 0.011$  &  $14.61 \pm 0.02$  &  $14.57 \pm 0.05$  &  12,26  &  &  $13.83 \pm 0.03$  &  $12.50 \pm 0.03$  &  16  &  &  --  &  --  &  $\cdots$  \\ 
 SDSSp J162414.37+002915.6  &  T6  &  &  $16.28 \pm 0.05$  &  $15.20 \pm 0.05$  &  $15.48 \pm 0.05$  &  $15.61 \pm 0.05$  &  1  &  &  $15.16 \pm 0.04$  &  $13.09 \pm 0.03$  &  16  &  &  $14.41 \pm 0.02$  &  $13.10 \pm 0.02$  &  24  \\ 
 2MASSI J1553022+153236  &  T7  &  &  $16.37 \pm 0.06$  &  $15.34 \pm 0.03$  &  $15.76 \pm 0.03$  &  $15.94 \pm 0.03$  &  2  &  &  $15.29 \pm 0.04$  &  $13.03 \pm 0.03$  &  16  &  &  $14.51 \pm 0.02$  &  $13.13 \pm 0.02$  &  24  \\ 
 WISE J031624.35+430709.1  &  T8  &  &  --  &  $19.47 \pm 0.04$  &  $19.70 \pm 0.09$  &  --  &  14  &  &  $17.79 \pm 0.22$  &  $14.64 \pm 0.05$  &  16  &  &  $16.64 \pm 0.04$  &  $14.58 \pm 0.02$  &  14  \\ 
 ULAS J130217.21+130851.2  &  T8.5  &  &  $19.12 \pm 0.03$  &  $18.11 \pm 0.04$  &  $18.60 \pm 0.06$  &  $18.28 \pm 0.03$  &  7  &  &  $17.69 \pm 0.23$  &  $14.87 \pm 0.07$  &  16  &  &  $16.510 \pm 0.010$  &  $14.92 \pm 0.02$  &  24  \\ 
\multicolumn{16}{l}{$\bullet$ Recovered Previously-Known Members} \\ 
 SIMP J013656.5+093347.3  &  T2.5  &  &  $14.392 \pm 0.003$  &  $13.252 \pm 0.002$  &  $12.809 \pm 0.002$  &  $12.585 \pm 0.002$  &  3,11  &  &  $11.94 \pm 0.02$  &  $10.96 \pm 0.02$  &  16  &  &  --  &  --  &  $\cdots$  \\ 
\cline{1-16} 
\multicolumn{16}{c}{Hyades} \\ 
\cline{1-16} 
\multicolumn{16}{l}{$\bullet$ New Candidate Members (with trigonometric parallax)} \\ 
 PSO J069.7303+04.3834  &  T2  &  &  --  &  $16.39 \pm 0.02$  &  $15.76 \pm 0.02$  &  $15.12 \pm 0.02$  &  25  &  &  $14.26 \pm 0.03$  &  $13.60 \pm 0.03$  &  16  &  &  --  &  --  &  $\cdots$  \\ 
 PSO J049.1159+26.8409  &  T2.5  &  &  $17.17 \pm 0.05$  &  $16.11 \pm 0.02$  &  $15.82 \pm 0.02$  &  $15.50 \pm 0.05$  &  19,26  &  &  $14.98 \pm 0.04$  &  $13.93 \pm 0.04$  &  16  &  &  --  &  --  &  $\cdots$  \\ 
 PSO J052.2746+13.3754  &  T3.5  &  &  $17.34 \pm 0.05$  &  $16.23 \pm 0.02$  &  $15.93 \pm 0.05$  &  $15.73 \pm 0.05$  &  25,26  &  &  $15.30 \pm 0.04$  &  $14.26 \pm 0.05$  &  16  &  &  --  &  --  &  $\cdots$  \\ 
\multicolumn{16}{l}{$\bullet$ New Candidate Members (with photometric parallax)} \\ 
 WISEPA J030724.57+290447.6  &  T6.5  &  &  $18.63 \pm 0.06$  &  $17.34 \pm 0.03$  &  $17.75 \pm 0.14$  &  $18.08 \pm 0.12$  &  3,11,26  &  &  $17.15 \pm 0.14$  &  $15.06 \pm 0.08$  &  16  &  &  $16.39 \pm 0.04$  &  $14.97 \pm 0.02$  &  9  \\ 
\multicolumn{16}{l}{$\bullet$ Recovered Previously-Known Candidate Members} \\ 
 CFHT$-$Hy$-$20  &  T2.5  &  &  $18.11 \pm 0.05$  &  $17.02 \pm 0.05$  &  $16.51 \pm 0.05$  &  $16.08 \pm 0.05$  &  4,21  &  &  $15.60 \pm 0.05$  &  $14.72 \pm 0.08$  &  16  &  &  --  &  --  &  $\cdots$  \\ 
\cline{1-16} 
\multicolumn{16}{c}{Ursa Major} \\ 
\cline{1-16} 
\multicolumn{16}{l}{$\bullet$ New Candidate Members (with trigonometric parallax)} \\ 
 SDSS J125011.65+392553.9  &  T4  &  &  --  &  $16.14 \pm 0.02$  &  $16.24 \pm 0.25$  &  $16.18 \pm 0.25$  &  25,26  &  &  $15.91 \pm 0.05$  &  $14.60 \pm 0.05$  &  16  &  &  --  &  --  &  $\cdots$  \\ 
\enddata 
\tablerefs{(1) \cite{1999ApJ...522L..61S}, (2) \cite{2004AJ....127.3553K}, (3) \cite{2007MNRAS.379.1599L}, (4) \cite{2008AandA...481..661B}, (5) \cite{2008ApJ...676.1281M}, (6) \cite{2008MNRAS.390..304P}, (7) \cite{2010MNRAS.406.1885B}, (8) \cite{2011AJ....141..203A}, (9) \cite{2011ApJS..197...19K}, (10) \cite{2012ApJ...753..156K}, (11) \cite{2012yCat.2314....0L}, (12) \cite{2013ApJ...777...84B}, (13) \cite{2013MNRAS.433..457B}, (14) \cite{2013ApJS..205....6M}, (15) \cite{2013Msngr.154...35M}, (16) \cite{2014yCat.2328....0C}, (17) \cite{2014ApJ...787....5N}, (18) \cite{2014ApJ...796...39T}, (19) \cite{2015ApJ...814..118B}, (20) \cite{2016yCat.2343....0E}, (21) \cite{2016ApJ...833...96L}, (22) \cite{2017AJ....154...10R}, (23) \cite{2018ApJS..236...28T}, (24) \cite{2019ApJS..240...19K}, (25) \cite{2020AJ....159..257B}, (26) \cite{2021AJ....161...42B}} 
\end{deluxetable} 
\end{longrotatetable}

\global\pdfpageattr\expandafter{\the\pdfpageattr/Rotate 90} 
\begin{longrotatetable} 
\begin{deluxetable}{lcccccccccccc} 
\tablewidth{0pc} 
\setlength{\tabcolsep}{0.05in} 
\tablecaption{Properties of T-Dwarf YMG Members and Candidates \label{tab:phys}} 
\tablehead{ \multicolumn{1}{l}{}  &  \multicolumn{1}{c}{}  &  \multicolumn{1}{c}{}  &  \multicolumn{1}{c}{}  &  \multicolumn{4}{c}{Physical Properties\tablenotemark{b}}  &  \multicolumn{1}{c}{}  &  \multicolumn{2}{c}{Peculiarity}  &  \multicolumn{1}{c}{}&  \multicolumn{1}{c}{}  \\ 
\cline{5-8} \cline{10-11} 
\multicolumn{1}{l}{Object}  &  \multicolumn{1}{c}{SpT}  &  \multicolumn{1}{c}{$\log{(L_{\rm bol}/L_{\odot})}$\tablenotemark{a,b}}  &  \multicolumn{1}{c}{}  &  \multicolumn{1}{c}{T$_{\rm eff}$}  &  \multicolumn{1}{c}{$\log{g}$}  &  \multicolumn{1}{c}{$R$}  &  \multicolumn{1}{c}{$M$}  &  \multicolumn{1}{c}{}  &  \multicolumn{1}{c}{Photometry\tablenotemark{c}}  &  \multicolumn{1}{c}{Spectroscopy\tablenotemark{d}}  &  \multicolumn{1}{c}{}  &  \multicolumn{1}{c}{Variability\tablenotemark{e}}  \\ 
\multicolumn{1}{l}{}  &  \multicolumn{1}{c}{}  &  \multicolumn{1}{c}{(dex)}  &  \multicolumn{1}{c}{}  &  \multicolumn{1}{c}{(K)}  &  \multicolumn{1}{c}{(dex)}  &  \multicolumn{1}{c}{(R$_{\rm Jup}$)}  &  \multicolumn{1}{c}{(M$_{\rm Jup}$)}  &  \multicolumn{1}{c}{}  &  \multicolumn{1}{c}{}  &  \multicolumn{1}{c}{}  &  \multicolumn{1}{c}{}  } 
\startdata 
\cline{1-13} 
\multicolumn{13}{c}{AB Doradus} \\ 
\cline{1-13} 
\multicolumn{13}{l}{$\bullet$ New Candidate Members (with trigonometric parallax)} \\ 
 WISE J163645.56$-$074325.1  &  T4.5  &  $-4.724 \pm 0.144$  &  &  $1090^{+94}_{-87}$  &  $4.36^{+0.08}_{-0.06}$  &  $1.19^{+0.02}_{-0.02}$  &  $13.1^{+2.2}_{-1.5}$  &  &  N  &  WCC\tablenotemark{g}  &  &    \\ 
 WISEPA J062720.07$-$111428.8  &  T6  &  $-5.045 \pm 0.103$  &  &  $910^{+53}_{-50}$  &  $4.29^{+0.04}_{-0.03}$  &  $1.178^{+0.011}_{-0.015}$  &  $11.1^{+0.9}_{-0.7}$  &  &  N  &  N  &  &    \\ 
 WISE J233226.49$-$432510.6  &  T9  &  $-6.242 \pm 0.040$  &  &  $454^{+11}_{-11}$  &  $3.82^{+0.08}_{-0.05}$  &  $1.193^{+0.006}_{-0.014}$  &  $3.8^{+0.7}_{-0.4}$  &  &  N  &  ?  &  &    \\ 
\multicolumn{13}{l}{$\bullet$ New Candidate Members (with photometric parallax)} \\ 
 ULAS J081918.58+210310.4  &  T6  &  $-5.013 \pm 0.158$  &  &  $926^{+84}_{-77}$  &  $4.30^{+0.05}_{-0.04}$  &  $1.179^{+0.013}_{-0.015}$  &  $11.2^{+1.2}_{-0.9}$  &  &  N  &  ?  &  &    \\ 
\multicolumn{13}{l}{$\bullet$ Recovered Previously-Known Members} \\ 
 2MASS J13243553+6358281  &  T2.5  &  $-4.720 \pm 0.100$  &  &  $1093^{+65}_{-63}$  &  $4.36^{+0.07}_{-0.05}$  &  $1.19^{+0.02}_{-0.02}$  &  $13.2^{+1.8}_{-1.3}$  &  &  Red  &  WCC\tablenotemark{f}  &  &  $3\pm0.3\%$  \\ 
 GU Psc b  &  T3.5  &  $-4.870 \pm 0.100$  &  &  $1002^{+59}_{-54}$  &  $4.32^{+0.05}_{-0.04}$  &  $1.185^{+0.012}_{-0.015}$  &  $11.9^{+1.3}_{-0.8}$  &  &  N  &  WCC\tablenotemark{f}  &  &  $4\pm1\%$  \\ 
 SDSSp J111010.01+011613.1  &  T5.5  &  $-4.970 \pm 0.020$  &  &  $948^{+12}_{-11}$  &  $4.30^{+0.04}_{-0.03}$  &  $1.184^{+0.009}_{-0.015}$  &  $11.3^{+0.9}_{-0.5}$  &  &  N  &  N  &  &  $<1.25\%$  \\ 
\cline{1-13} 
\multicolumn{13}{c}{Argus} \\ 
\cline{1-13} 
\multicolumn{13}{l}{$\bullet$ New Candidate Members (with trigonometric parallax)} \\ 
 SDSS J152103.24+013142.7  &  T3  &  $-4.666 \pm 0.149$  &  &  $1083^{+91}_{-86}$  &  $4.11^{+0.04}_{-0.04}$  &  $1.284^{+0.015}_{-0.010}$  &  $8.5^{+0.9}_{-0.9}$  &  &  N  &  WCC  &  &    \\ 
 2MASS J00132229$-$1143006  &  T4  &  $-4.740 \pm 0.121$  &  &  $1039^{+71}_{-67}$  &  $4.09^{+0.03}_{-0.04}$  &  $1.279^{+0.009}_{-0.008}$  &  $8.1^{+0.7}_{-0.7}$  &  &  N  &  WCC\tablenotemark{f,g}  &  &  $4.6\pm0.2\%$  \\ 
 SDSS J020742.48+000056.2  &  T4.5  &  $-4.768 \pm 0.134$  &  &  $1024^{+78}_{-73}$  &  $4.08^{+0.04}_{-0.04}$  &  $1.278^{+0.010}_{-0.008}$  &  $7.9^{+0.8}_{-0.8}$  &  &  N  &  N  &  &    \\ 
 WISE J024124.73$-$365328.0  &  T7  &  $-5.312 \pm 0.075$  &  &  $754^{+33}_{-31}$  &  $3.90^{+0.03}_{-0.03}$  &  $1.259^{+0.005}_{-0.005}$  &  $5.1^{+0.4}_{-0.4}$  &  &  N  &  N  &  &    \\ 
\multicolumn{13}{l}{$\bullet$ New Candidate Members (with photometric parallax)} \\ 
 ULAS J004757.41+154641.4  &  T2  &  $-4.710 \pm 0.161$  &  &  $1057^{+96}_{-90}$  &  $4.10^{+0.04}_{-0.05}$  &  $1.281^{+0.013}_{-0.010}$  &  $8.3^{+0.9}_{-1.0}$  &  &  Red  &  SCC\tablenotemark{g}  &  &    \\ 
 PSO J168.1800$-$27.2264  &  T2.5  &  $-4.754 \pm 0.108$  &  &  $1032^{+63}_{-60}$  &  $4.08^{+0.03}_{-0.04}$  &  $1.278^{+0.009}_{-0.008}$  &  $8.0^{+0.7}_{-0.7}$  &  &  N  &  SCC\tablenotemark{f}  &  &    \\ 
 ULAS J154701.84+005320.3  &  T5.5  &  $-5.140 \pm 0.170$  &  &  $831^{+83}_{-76}$  &  $3.96^{+0.06}_{-0.07}$  &  $1.264^{+0.007}_{-0.006}$  &  $5.9^{+0.9}_{-0.9}$  &  &  N  &  ?  &  &    \\ 
 ULAS J120744.65+133902.7  &  T6  &  $-5.355 \pm 0.159$  &  &  $736^{+69}_{-63}$  &  $3.88^{+0.06}_{-0.07}$  &  $1.258^{+0.006}_{-0.005}$  &  $4.9^{+0.8}_{-0.7}$  &  &  N  &  ?  &  &    \\ 
 ULAS J075829.83+222526.7  &  T6.5  &  $-5.375 \pm 0.163$  &  &  $727^{+70}_{-64}$  &  $3.87^{+0.06}_{-0.07}$  &  $1.258^{+0.006}_{-0.005}$  &  $4.8^{+0.8}_{-0.7}$  &  &  N  &  ?  &  &    \\ 
\cline{1-13} 
\multicolumn{13}{c}{$\beta$ Pictoris} \\ 
\cline{1-13} 
\multicolumn{13}{l}{$\bullet$ New Candidate Members (with trigonometric parallax)} \\ 
 WISEPA J081958.05$-$033529.0  &  T4  &  $-4.769 \pm 0.078$  &  &  $1004^{+44}_{-42}$  &  $3.91^{+0.04}_{-0.04}$  &  $1.325^{+0.011}_{-0.010}$  &  $5.7^{+0.5}_{-0.5}$  &  &  N  &  WCC  &  &    \\ 
 CFBDS J232304.41$-$015232.3  &  T6  &  $-4.979 \pm 0.147$  &  &  $894^{+75}_{-70}$  &  $3.84^{+0.06}_{-0.06}$  &  $1.311^{+0.013}_{-0.010}$  &  $4.8^{+0.7}_{-0.7}$  &  &  N  &  ?  &  &    \\ 
 WISEPC J225540.74$-$311841.8  &  T8  &  $-5.787 \pm 0.057$  &  &  $577^{+16}_{-14}$  &  $3.54^{+0.03}_{-0.02}$  &  $1.274^{+0.005}_{-0.005}$  &  $2.3^{+0.2}_{-0.1}$  &  &  Red  &  N  &  &    \\ 
\multicolumn{13}{l}{$\bullet$ Recovered Previously-Known Members} \\ 
 51 Eri b  &  T6.5  &  $-5.870 \pm 0.150$  &  &  $588^{+35}_{-25}$  &  $3.55^{+0.05}_{-0.03}$  &  $1.276^{+0.007}_{-0.006}$  &  $2.3^{+0.3}_{-0.2}$  &  &  Red+Faint  &  N  &  &    \\ 
\cline{1-13} 
\multicolumn{13}{c}{Carina-Near} \\ 
\cline{1-13} 
\multicolumn{13}{l}{$\bullet$ {\bf Newly Confirmed Member}} \\ 
 {\bf 2MASS J21392676+0220226}  &  T1.5  &  $-4.710 \pm 0.056$  &  &  $1111^{+37}_{-42}$  &  $4.42^{+0.12}_{-0.06}$  &  $1.17^{+0.02}_{-0.04}$  &  $14.6^{+3.2}_{-1.6}$  &  &  Red+Faint  &  WCC\tablenotemark{f,g}  &  &  $26\pm1\%$  \\ 
\multicolumn{13}{l}{$\bullet$ New Candidate Members (with trigonometric parallax)} \\ 
 ULAS J131610.13+031205.5  &  T3  &  $-4.431 \pm 0.131$  &  &  $1284^{+113}_{-77}$  &  $4.57^{+0.18}_{-0.16}$  &  $1.17^{+0.05}_{-0.04}$  &  $20.3^{+9.1}_{-5.0}$  &  &  Red  &  SCC\tablenotemark{g}  &  &    \\ 
 PSO J004.6359+56.8370  &  T4.5  &  $-4.956 \pm 0.093$  &  &  $960^{+51}_{-47}$  &  $4.35^{+0.06}_{-0.05}$  &  $1.170^{+0.014}_{-0.021}$  &  $12.3^{+1.4}_{-1.0}$  &  &  N  &  N  &  &    \\ 
 WISE J223617.59+510551.9  &  T5  &  $-4.975 \pm 0.069$  &  &  $950^{+41}_{-35}$  &  $4.34^{+0.06}_{-0.04}$  &  $1.17^{+0.02}_{-0.02}$  &  $12.1^{+1.3}_{-0.9}$  &  &  N  &  N  &  &    \\ 
 SDSSp J162414.37+002915.6  &  T6  &  $-5.238 \pm 0.062$  &  &  $820^{+30}_{-29}$  &  $4.31^{+0.04}_{-0.05}$  &  $1.156^{+0.016}_{-0.013}$  &  $11.0^{+0.8}_{-1.0}$  &  &  N  &  N  &  &    \\ 
 2MASSI J1553022+153236  &  T7  &  $-5.021 \pm 0.093$  &  &  $927^{+50}_{-47}$  &  $4.34^{+0.06}_{-0.04}$  &  $1.17^{+0.02}_{-0.02}$  &  $12.0^{+1.3}_{-1.0}$  &  &  N  &  N  &  &    \\ 
 WISE J031624.35+430709.1  &  T8  &  $-6.180 \pm 0.038$  &  &  $473^{+11}_{-11}$  &  $3.93^{+0.07}_{-0.08}$  &  $1.178^{+0.015}_{-0.013}$  &  $4.8^{+0.7}_{-0.7}$  &  &  Faint  &  ?  &  &    \\ 
 ULAS J130217.21+130851.2  &  T8.5  &  $-6.035 \pm 0.077$  &  &  $514^{+24}_{-23}$  &  $4.00^{+0.08}_{-0.09}$  &  $1.175^{+0.017}_{-0.015}$  &  $5.6^{+0.9}_{-0.9}$  &  &  Red  &  ?  &  &    \\ 
\multicolumn{13}{l}{$\bullet$ Recovered Previously-Known Members} \\ 
 SIMP J013656.5+093347.3  &  T2.5  &  $-4.688 \pm 0.005$  &  &  $1126^{+16}_{-15}$  &  $4.46^{+0.09}_{-0.08}$  &  $1.16^{+0.03}_{-0.03}$  &  $15.6^{+2.4}_{-2.1}$  &  &  N  &  WCC  &  &  $\approx 5\%$  \\ 
\cline{1-13} 
\multicolumn{13}{c}{Hyades} \\ 
\cline{1-13} 
\multicolumn{13}{l}{$\bullet$ New Candidate Members (with trigonometric parallax)} \\ 
 PSO J069.7303+04.3834  &  T2  &  $-4.390 \pm 0.160$  &  &  $1445^{+147}_{-135}$  &  $5.10^{+0.08}_{-0.13}$  &  $0.99^{+0.02}_{-0.02}$  &  $50.0^{+7.7}_{-11.9}$  &  &  Red  &  SCC\tablenotemark{g}  &  &    \\ 
 PSO J049.1159+26.8409  &  T2.5  &  $-4.521 \pm 0.096$  &  &  $1331^{+83}_{-72}$  &  $5.00^{+0.10}_{-0.09}$  &  $1.00^{+0.02}_{-0.02}$  &  $40.6^{+8.5}_{-7.1}$  &  &  Bright  &  SCC\tablenotemark{f,g}  &  &    \\ 
 PSO J052.2746+13.3754  &  T3.5  &  $-4.770 \pm 0.136$  &  &  $1154^{+94}_{-88}$  &  $4.86^{+0.07}_{-0.05}$  &  $1.003^{+0.015}_{-0.014}$  &  $29.6^{+4.3}_{-2.9}$  &  &  N  &  N  &  &    \\ 
\multicolumn{13}{l}{$\bullet$ New Candidate Members (with photometric parallax)} \\ 
 WISEPA J030724.57+290447.6  &  T6.5  &  $-5.480 \pm 0.123$  &  &  $755^{+59}_{-55}$  &  $4.63^{+0.06}_{-0.09}$  &  $1.03^{+0.02}_{-0.02}$  &  $18.4^{+2.3}_{-2.7}$  &  &  N  &  ?  &  &    \\ 
\multicolumn{13}{l}{$\bullet$ Recovered Previously-Known Candidate Members} \\ 
 CFHT$-$Hy$-$20  &  T2.5  &  $-4.663 \pm 0.071$  &  &  $1220^{+41}_{-40}$  &  $4.90^{+0.05}_{-0.05}$  &  $1.003^{+0.015}_{-0.014}$  &  $32.0^{+3.1}_{-2.6}$  &  &  N  &  WCC  &  &    \\ 
\cline{1-13} 
\multicolumn{13}{c}{Ursa Major} \\ 
\cline{1-13} 
\multicolumn{13}{l}{$\bullet$ New Candidate Members (with trigonometric parallax)} \\ 
 SDSS J125011.65+392553.9  &  T4  &  $-4.909 \pm 0.128$  &  &  $1023^{+88}_{-79}$  &  $4.65^{+0.14}_{-0.29}$  &  $1.07^{+0.10}_{-0.05}$  &  $20.6^{+5.2}_{-8.0}$  &  &  N  &  ?  &  &    \\ 
\enddata 
\tablenotetext{a}{Bolometric luminosities for our candidate members with photometric parallaxes should be used with caution.} 
\tablenotetext{b}{Bolometric luminosities and physical properties of previously-known YMG members and candidate members are from Table~4 of \cite{2020ApJ...891..171Z}.} 
\tablenotetext{c}{Objects with ``Red'' have much redder $J-K$ colors than field dwarfs with similar spectral types, and those with ``Faint'' or ``Bright'' have much fainter or brighter $J$-band absolute magnitudes than the field sequence (see Figure~\ref{fig:cmd}).} 
\tablenotetext{d}{Objects with ``SCC'' and ``WCC'' are strong and weak composite candidates, respectively, based on the \cite{2010ApJ...710.1142B} and \cite{2015AJ....150..163B} criteria (see Section~\ref{subsec:spec}). We use ``?'' for objects whose spectra have only partial wavelength coverage in the near-infrared and/or have not been vetted for spectral peculiarity by previous work. We use ``N'' for objects with normal spectra.} 
\tablenotetext{e}{We provide peak-to-peak amplitudes in $J$ band for variable ultracool dwarfs 2MASS~J$00132229-1143006$ \citep{2019AandA...629A.145E}, 2MASS~J$21392676+0220226$ \citep{2012ApJ...750..105R}, GU~Psc~b \citep[][]{2017AJ....154..138N}, and SIMP~J$013656.5+093347.3$ \citep{2009ApJ...701.1534A, 2014ApJ...793...75R, 2017ApJ...842...78V}, and in {\it Spitzer}/IRAC [4.5] band for SDSSp~J$111010.01+011613.1$ \citep{2018MNRAS.474.1041V} and 2MASS~J$13243559+6358284$ \citep{2015ApJ...799..154M}. } 
\tablenotetext{f}{Potential binarity of our 4 YMG candidates and 2 previously-known YMG members have been previously noted by using the same quantitative spectral indices as in this work: 2MASS~J$21392676+0220226$ and 2MASS~J$13243553+6358281$ by \cite{2010ApJ...710.1142B}, GU~Psc~b by \cite{2014ApJ...787....5N}, PSO~J$049.1159+26.8409$ and PSO~J$168.1800-27.2264$ by \cite{2015ApJ...814..118B}, and 2MASS~J$00132229-1143006$ by \cite{2017AJ....154..112K}.}
\tablenotetext{g}{These 7 objects have peculiar spectra indicative of either atmospheric variability or unresolved binarity (Section~\ref{subsec:spec}).} 
\end{deluxetable} 
\end{longrotatetable}

\global\pdfpageattr\expandafter{\the\pdfpageattr/Rotate 90} 
\begin{longrotatetable} 
\begin{deluxetable}{lccccccccccccccccc} 
\tablewidth{0pc} 
\setlength{\tabcolsep}{0.05in} 
\tablecaption{T7$-$Y1 Benchmarks: Spectral Type, Photometry, Parallax, and Age \label{tab:lateT_basic}} 
\tablehead{ \multicolumn{1}{l}{}  &  \multicolumn{1}{c}{}  &  \multicolumn{1}{c}{}  &  \multicolumn{4}{c}{MKO Photometry}  &  \multicolumn{1}{c}{}  &  \multicolumn{1}{c}{}  &  \multicolumn{1}{c}{}  &  \multicolumn{1}{c}{}  &  \multicolumn{1}{c}{}  &  \multicolumn{1}{c}{}  &  \multicolumn{5}{c}{References}  \\ 
\cline{4-7}  \cline{14-18}  
\multicolumn{1}{l}{Object}  &  \multicolumn{1}{c}{SpT}  &  \multicolumn{1}{c}{}  &  \multicolumn{1}{c}{$Y_{\rm MKO}$}  &  \multicolumn{1}{c}{$J_{\rm MKO}$}  &  \multicolumn{1}{c}{$H_{\rm MKO}$}  &  \multicolumn{1}{c}{$K_{\rm MKO}$}  &  \multicolumn{1}{c}{}  &  \multicolumn{1}{c}{Parallax}  &  \multicolumn{1}{c}{Age}  &  \multicolumn{1}{c}{Primary SpT}  &  \multicolumn{1}{c}{Separation}  &  \multicolumn{1}{c}{}  &  \multicolumn{1}{c}{Discovery}  &  \multicolumn{1}{c}{SpT}  &  \multicolumn{1}{c}{Phot.}  &  \multicolumn{1}{c}{$\pi$}  &  \multicolumn{1}{c}{Age}  \\ 
\multicolumn{1}{l}{}  &  \multicolumn{1}{c}{}  &  \multicolumn{1}{c}{}  &  \multicolumn{1}{c}{(mag)}  &  \multicolumn{1}{c}{(mag)}  &  \multicolumn{1}{c}{(mag)}  &  \multicolumn{1}{c}{(mag)}  &  \multicolumn{1}{c}{}  &  \multicolumn{1}{c}{(mas)}  &  \multicolumn{1}{c}{(Gyr)}  &  \multicolumn{1}{c}{}  &  \multicolumn{1}{c}{($''$)}  &  \multicolumn{1}{c}{}  &  \multicolumn{1}{c}{}  &  \multicolumn{1}{c}{}  &  \multicolumn{1}{c}{}  &  \multicolumn{1}{c}{}  &  \multicolumn{1}{c}{}   } 
\startdata 
 Gl~229B  &  T7 pec  &  &  $15.17 \pm 0.10$  &  $14.01 \pm 0.05$  &  $14.36 \pm 0.05$  &  $14.36 \pm 0.05$  &  &  $173.70 \pm 0.05$  &  $8.6^{+1.4}_{-0.6}$  &  M1V  &  $7.8$  &   &  1  &  4  &  3  &  30,34  &  37  \\ 
 HD~3651B  &  T7.5  &  &  $17.12 \pm 0.06$  &  $16.16 \pm 0.03$  &  $16.68 \pm 0.04$  &  $16.87 \pm 0.05$  &  &  $89.79 \pm 0.06$  &  $4.5-8.3$  &  K0V  &  $42.9$  &   &  5  &  6  &  6,41  &  30,34  &  40  \\ 
 ULAS~J141623.94+134836.3  &  sd~T7.5  &  &  $18.16 \pm 0.03$  &  $17.26 \pm 0.02$  &  $17.58 \pm 0.03$  &  $18.42 \pm 0.09$  &  &  $107.56 \pm 0.30$  &  $0.5-10$  &  sd~L7  &  $9.8$  &   &  11,15  &  10  &  19,41  &  30,34  &  28  \\ 
 GJ~570D  &  T7.5  &  &  $15.78 \pm 0.10$  &  $14.82 \pm 0.05$  &  $15.28 \pm 0.05$  &  $15.52 \pm 0.05$  &  &  $170.01 \pm 0.09$  &  $1.4-5.2$  &  K4V + M1.5V + M3V  &  $258.3$  &   &  2  &  4  &  14  &  30,34  &  40  \\ 
 ULAS~J095047.28+011734.3  &  T8  &  &  $18.90 \pm 0.03$  &  $18.02 \pm 0.03$  &  $18.40 \pm 0.03$  &  $18.85 \pm 0.07$  &  &  $50.80 \pm 0.08$  &  $>3.5$  &  M4V  &  $52.0$  &   &  22  &  26  &  22  &  30,34  &  22  \\ 
 Ross~458C  &  T8  &  &  $17.72 \pm 0.03$  &  $16.69 \pm 0.01$  &  $17.01 \pm 0.04$  &  $16.90 \pm 0.06$  &  &  $86.86 \pm 0.15$  &  $0.15-0.8$  &  M0.5V + M7V  &  $103.0$  &   &  13  &  16  &  19  &  30,34  &  12  \\ 
 BD~+01$^{\circ}$~2920B  &  T8  &  &  $19.51 \pm 0.14$  &  $18.55 \pm 0.03$  &  $18.96 \pm 0.07$  &  $19.89 \pm 0.33$  &  &  $58.20 \pm 0.50$  &  $2.3-14.4$  &  G1V  &  $153.0$  &   &  21  &  26  &  19,21,26  &  7  &  21  \\ 
 WISEU~J005559.88+594745.0  &  T8  &  &  --  &  $17.90 \pm 0.05$  &  --  &  --  &  &  $43.78 \pm 0.07$  &  $10\pm3$  &  DC  &  $17.6$  &   &  39  &  39  &  39  &  30,34  &  39  \\ 
 WISEU~J215018.99$-$752054.6  &  T8  &  &  $18.53 \pm 0.13$  &  $18.10 \pm 0.10$  &  --  &  --  &  &  $41.36 \pm 0.28$  &  $0.5-10$  &  L1  &  $14.1$  &   &  38  &  38  &  39  &  30,34  &  38  \\ 
 Wolf~1130C  &  sd~T8  &  &  --  &  $19.64 \pm 0.09$  &  $19.57 \pm 0.08$  &  --  &  &  $60.39 \pm 0.03$  &  $>10$  &  sdM1 + WD  &  $188.5$  &   &  25  &  25  &  25  &  30,34  &  35  \\ 
 WISE~J111838.70+312537.9  &  T8.5  &  &  $19.18 \pm 0.12$  &  $17.79 \pm 0.05$  &  $18.15 \pm 0.06$  &  $18.75 \pm 0.15$  &  &  $114.50 \pm 0.40$  &  $>2$  &  F8.5V + G2V\tablenotemark{a}  &  $510.0$  &   &  27  &  27  &  27  &  30,34  &  27  \\ 
 Wolf~940B  &  T8.5  &  &  $18.97 \pm 0.03$  &  $18.16 \pm 0.02$  &  $18.77 \pm 0.03$  &  $18.85 \pm 0.05$  &  &  $80.77 \pm 0.11$  &  $3.5-6.0$  &  dM4  &  $32.0$  &   &  8  &  8  &  8  &  30,34  &  8  \\ 
 Gl~758B  &  T5$-$T8  &  &  --  &  $18.57 \pm 0.20$  &  $19.15 \pm 0.20$  &  --  &  &  $64.06 \pm 0.02$  &  $8.8\pm0.9$  &  K0V  &  $1.9$  &   &  9  &  33  &  17  &  30,34  &  42  \\ 
 GJ~504~b  &  late-T  &  &  --  &  $19.76 \pm 0.10$  &  $19.99 \pm 0.10$  &  $19.38 \pm 0.11$  &  &  $57.02 \pm 0.25$  &  $0.1-6.5$  &  G0V  &  $2.5$  &   &  24  &  24  &  23  &  30,34  &  31  \\ 
 WD~0806$-$661B  &  Y1  &  &  --  &  $25.00 \pm 0.10$  &  $25.29 \pm 0.14$  &  --  &  &  $51.93 \pm 0.02$  &  $2\pm0.5$  &  DQ  &  $130.2$  &   &  18  &  36  &  29,32  &  30,34  &  20  \\ 
\enddata 
\tablenotetext{a}{The two primary stars of WISE~J$111838.70+312537.9$ form a gravitationally bound binary system \citep{1804RSPT...94..353H} and are both spectroscopic binaries \citep[e.g.,][]{1967AN....289..269H}.} 
\tablerefs{(1) \cite{1995Natur.378..463N}, (2) \cite{2000ApJ...531L..57B}, (3) \cite{2002MNRAS.332...78L}, (4) \cite{2006ApJ...637.1067B}, (5) \cite{2006MNRAS.373L..31M}, (6) \cite{2007ApJ...654..570L}, (7) \cite{2007AandA...474..653V}, (8) \cite{2009MNRAS.395.1237B}, (9) \cite{2009ApJ...707L.123T}, (10) \cite{2010AJ....139.2448B}, (11) \cite{2010MNRAS.404.1952B}, (12) \cite{2010ApJ...725.1405B}, (13) \cite{2010MNRAS.405.1140G}, (14) \cite{2010ApJ...710.1627L}, (15) \cite{2010AandA...510L...8S}, (16) \cite{2011ApJ...743...50C}, (17) \cite{2011ApJ...728...85J}, (18) \cite{2011ApJ...730L...9L}, (19) \cite{2012yCat.2314....0L}, (20) \cite{2012ApJ...744..135L}, (21) \cite{2012MNRAS.422.1922P}, (22) \cite{2013MNRAS.433..457B}, (23) \cite{2013ApJ...778L...4J}, (24) \cite{2013ApJ...774...11K}, (25) \cite{2013ApJ...777...36M}, (26) \cite{2013ApJS..205....6M}, (27) \cite{2013AJ....145...84W}, (28) \cite{2015ApJ...810..158F}, (29) \cite{2015ApJ...799...37L}, (30) \cite{2016AandA...595A...1G}, (31) \cite{2016ApJ...817..166S}, (32) \cite{2017ApJ...842..118L}, (33) \cite{2017ApJ...838...64N}, (34) \cite{2018AandA...616A...1G}, (35) \cite{2018ApJ...854..145M}, (36) \cite{2019ApJS..240...19K}, (37) \cite{2020AJ....160..196B}, (38) \cite{2020ApJ...889..176F}, (39) \cite{2020ApJ...899..123M}, (40) \cite{2020arXiv201112294Z}, (41) \cite{2021AJ....161...42B}, (42) This Work} 
\end{deluxetable} 
\end{longrotatetable}

\global\pdfpageattr\expandafter{\the\pdfpageattr/Rotate 90} 
\begin{longrotatetable} 
\begin{deluxetable}{lccccccccccc} 
\tablewidth{0pc} 
\tablecaption{T7$-$Y1 Benchmarks: Bolometric Luminosity, Effective Temperature, Surface Gravity, Radius, and Mass \label{tab:lateT_phys}} 
\tablehead{ \multicolumn{1}{l}{}  &  \multicolumn{1}{c}{}  &  \multicolumn{1}{c}{}  &  \multicolumn{1}{c}{}  &  \multicolumn{4}{c}{Physical Properties\tablenotemark{b}}  &  \multicolumn{1}{c}{}  &  \multicolumn{1}{c}{}  &  \multicolumn{2}{c}{References}  \\ 
\cline{5-9} \cline{11-12}  
\multicolumn{1}{l}{Object}  &  \multicolumn{1}{c}{SpT}  &  \multicolumn{1}{c}{$\log{(L_{\rm bol}/L_{\rm \odot})}$}  &  \multicolumn{1}{c}{}  &  \multicolumn{1}{c}{$T_{\rm eff}$}  &  \multicolumn{1}{c}{$\log{g}$}  &  \multicolumn{1}{c}{$R$}  &  \multicolumn{1}{c}{$M$}  &  \multicolumn{1}{c}{}  &  \multicolumn{1}{c}{}  &  \multicolumn{1}{c}{$L_{\rm bol}$}  &  \multicolumn{1}{c}{Phys.}  \\ 
\multicolumn{1}{l}{}  &  \multicolumn{1}{c}{}  &  \multicolumn{1}{c}{(dex)}  &  \multicolumn{1}{c}{}  &  \multicolumn{1}{c}{(K)}  &  \multicolumn{1}{c}{(dex)}  &  \multicolumn{1}{c}{($R_{\rm Jup}$)}  &  \multicolumn{1}{c}{($M_{\rm Jup}$)}  &  \multicolumn{1}{c}{}  &  \multicolumn{1}{c}{}  &  \multicolumn{1}{c}{}  &  \multicolumn{1}{c}{}     } 
\startdata 
 Gl~229B  &  T7 pec  &  $-5.208 \pm 0.007$  &  &  $1011^{+6}_{-5}$  &  $5.40^{+0.02}_{-0.01}$  &  $0.79^{+0.00}_{-0.01}$  &  $70.0^{+5.0}_{-5.0}$\tablenotemark{a}  &  &  &  4  &  8  \\ 
 HD~3651B  &  T7.5  &  $-5.57 \pm 0.03$  &  &  $802^{+17}_{-17}$  &  $5.24^{+0.05}_{-0.07}$  &  $0.83^{+0.02}_{-0.02}$  &  $47.8^{+4.0}_{-4.6}$  &  &  &  10  &  11  \\ 
 ULAS~J141623.94+134836.3  &  sd~T7.5  &  $-5.79 \pm 0.01$  &  &  $691^{+19}_{-36}$  &  $5.09^{+0.15}_{-0.29}$  &  $0.87^{+0.10}_{-0.05}$  &  $37.4^{+9.4}_{-14.0}$  &  &  &  4  &  11  \\ 
 GJ~570D  &  T7.5  &  $-5.54 \pm 0.03$  &  &  $786^{+21}_{-23}$  &  $5.06^{+0.10}_{-0.15}$  &  $0.89^{+0.05}_{-0.03}$  &  $36.2^{+5.8}_{-7.6}$  &  &  &  10  &  11  \\ 
 ULAS~J095047.28+011734.3  &  T8  &  $-5.63 \pm 0.07$  &  &  $774^{+36}_{-36}$  &  $5.23^{+0.08}_{-0.12}$  &  $0.83^{+0.04}_{-0.03}$  &  $47.0^{+6.4}_{-7.8}$  &  &  &  3  &  11  \\ 
 Ross~458C  &  T8  &  $-5.60^{+0.03}_{-0.04}$  &  &  $683^{+18}_{-19}$  &  $4.41^{+0.13}_{-0.15}$  &  $1.10^{+0.04}_{-0.04}$  &  $12.5^{+3.2}_{-2.8}$  &  &  &  10  &  11  \\ 
 BD~+01$^{\circ}$~2920B  &  T8  &  $-5.83 \pm 0.05$  &  &  $677^{+27}_{-27}$  &  $5.12^{+0.11}_{-0.17}$  &  $0.85^{+0.06}_{-0.03}$  &  $39.0^{+7.1}_{-9.1}$  &  &  &  2  &  11  \\ 
 WISEU~J005559.88+594745.0\tablenotemark{b}  &  T8  &  $-5.712 \pm 0.008$  &  &  $753$  &  $5.317$  &  $0.7955$  &  $53.1$  &  &  &  11  &  11  \\ 
 WISEU~J215018.99$-$752054.6  &  T8  &  $-5.64\pm0.02$  &  &  $758^{+23}_{-39}$  &  $5.15^{+0.14}_{-0.29}$  &  $0.85^{+0.10}_{-0.05}$  &  $41.7^{+10.1}_{-15.1}$  &  &  &  9  &  11  \\ 
 Wolf~1130C\tablenotemark{b}  &  sd~T8  &  $-5.949 \pm 0.007$  &  &  $647$  &  $5.219$  &  $0.8189$  &  $44.9$  &  &  &  11  &  11  \\ 
 WISE~J111838.70+312537.9  &  T8.5  &  $-6.063 \pm 0.008$  &  &  $584^{+13}_{-19}$  &  $5.01^{+0.12}_{-0.19}$  &  $0.88^{+0.06}_{-0.04}$  &  $31.9^{+6.4}_{-8.3}$  &  &  &  11  &  11  \\ 
 Wolf~940B  &  T8.5  &  $-6.07 \pm 0.04$  &  &  $574^{+16}_{-16}$  &  $4.93^{+0.05}_{-0.06}$  &  $0.91^{+0.02}_{-0.02}$  &  $28.2^{+2.6}_{-2.7}$  &  &  &  1  &  11  \\ 
 Gl~758B  &  T5$-$T8  &  $-6.07 \pm 0.03$  &  &  $594^{+10}_{-10}$  &  $5.12^{+0.03}_{-0.03}$  &  $0.85^{+0.01}_{-0.01}$  &  $37.9^{+1.5}_{-1.5}$\tablenotemark{a}  &  &  &  6  &  7,11  \\ 
 GJ~504~b  &  late-T  &  $-6.13 \pm 0.03$  &  &  $540^{+18}_{-28}$  &  $4.79^{+0.15}_{-0.36}$  &  $0.95^{+0.11}_{-0.05}$  &  $22.3^{+6.3}_{-10.2}$  &  &  &  5  &  11  \\ 
 WD~0806$-$661B  &  Y1  &  $-6.983 \pm 0.017$  &  &  $328^{+4}_{-4}$  &  $4.22^{+0.07}_{-0.09}$  &  $1.07^{+0.02}_{-0.02}$  &  $7.8^{+1.0}_{-1.2}$  &  &  &  11  &  11  \\ 
\enddata 
\tablenotetext{a}{Directly measured dynamical masses of Gl~229B and Gl~758B are listed in the table and their estimated masses from the \cite{2008ApJ...689.1327S} hybrid evolutionary models (using their dynamical masses as the prior and their measured $L_{\rm bol}$) are $63.1^{+1.7}_{-0.9}$~M$_{\rm Jup}$ \citep{2020AJ....160..196B} and $38.1 \pm 1.7$~M$_{\rm Jup}$ (Section~\ref{subsec:physical}), respectively.} 
\tablenotetext{b}{Physical properties of WISEU~J005559.88+594745.0 and Wolf~1130C are derived by assuming an age of 10~Gyr. Only median values of these two objects' properties are shown here, with uncertainties of 4~K in $T_{\rm eff}$, 0.003~dex in $\log{g}$, $8 \times 10^{-4}$~R$_{\rm Jup}$ in $R$, and $0.3$~M$_{\rm Jup}$ in $M$ for both objects.} 
\tablerefs{(1) \cite{2009MNRAS.395.1237B}, (2) \cite{2012MNRAS.422.1922P}, (3) \cite{2013MNRAS.433..457B}, (4) \cite{2015ApJ...810..158F}, (5) \cite{2016ApJ...817..166S}, (6) \cite{2018AJ....155..159B}, (7) \cite{2019AJ....158..140B}, (8) \cite{2020AJ....160..196B}, (9) \cite{2020ApJ...889..176F}, (10) \cite{2020arXiv201112294Z}, (11) This Work} 
\end{deluxetable} 
\end{longrotatetable}

\vfill
\eject
\end{document}